\documentclass[review,number,sort&compress]{elsarticle}

\usepackage{graphicx}
\usepackage{amssymb}
\usepackage{amsmath}
\usepackage{bm}
\usepackage{graphicx}
\usepackage{mathrsfs}
\usepackage{multirow}
\usepackage{subfig}

\usepackage{url}

\begin{document}

\begin{frontmatter}

\title{High accuracy position response calibration method for a micro-channel plate ion detector}

\author[UW,ANLHEP]{R.~Hong\corref{cor_auth}}\ead{hongran@uw.edu}
\author[ANL]{A.~Leredde}
\author[UW]{Y.~Bagdasarova}
\author[LPC]{X.~Fl\'echard}
\author[UW]{A.~Garc\'{\i}a}
\author[ANL]{P.~M\"uller}
\author[PSI]{A.~Knecht}
\author[LPC]{E.~Li\'enard}
\author[UMD]{M.~Kossin}
\author[UW]{M.G.~Sternberg}
\author[UW]{H.E.~Swanson}
\author[UW]{D.W.~Zumwalt}

\address[UW]{Department of Physics and Center for Experimental Nuclear Physics and Astrophysics, University of Washington, Seattle, WA 98195, USA}
\address[ANLHEP]{High Energy Physics Division, Argonne National Laboratory, 9700 S Cass Ave, Argonne, Illinois 60439, USA}
\address[ANL]{Physics Division, Argonne National Laboratory, 9700 S Cass Ave, Argonne, Illinois 60439, USA}
\address[LPC]{Normandie Univ, ENSICAEN, UNICAEN, CNRS/IN2P3, LPC Caen, 14000 Caen, France}
\address[PSI]{Paul Scherrer Institut,5232 Villigen PSI,Switzerland}
%\address[NSCL]{National Superconducting Cyclotron Laboratory and Department of Physics and Astronomy,Michigan State University, East Lansing 48824 MI, USA}
\address[UMD]{Department of Physics, University of Maryland, College Park, MD 20742}

\cortext[cor_auth]{Corresponding author}

\begin{abstract}
We have developed a position response calibration method for a micro-channel plate (MCP) detector with a delay-line anode position readout scheme. Using an in situ calibration mask, an accuracy of 8~$\mu$m and a resolution of 85~$\mu$m (FWHM) have been achieved for MeV-scale $\alpha$ particles and ions with energies of $\sim$10~keV. At this level of accuracy, the difference between the MCP position responses to high-energy $\alpha$ particles and low-energy ions is significant. The improved performance of the MCP detector can find applications in many fields of AMO and nuclear physics. In our case, it helps reducing systematic uncertainties in a high-precision nuclear $\beta$-decay experiment.

\end{abstract}

\end{frontmatter}

\section{Introduction}
\label{Sec_Introduction}

For more than two decades micro-channel plates (MCPs) have been widely used for ion and electron detection in experiments dedicated to the study of ion, atom and molecule collisions and photoionizations using the so called ``cold target recoil ion momentum spectroscopy'' (COLTRIMS) technique \cite{Dorner_2000}. More recently, such detectors found new applications in nuclear physics with the precise measurement of atomic masses of short lived nuclides \cite{Eliseev_2013} and high-precision nuclear $\beta$-decay experiments \cite{Gorelov.142501,Vetter_2008,Flechard_2011,VanGorp_2014}. Many of these experiments demand high accuracy and resolution for both the timing and position responses of the MCP. For example, in the measurement of the $\beta-\nu$ angular correlation in the $^{6}$He decay at University of Washington \cite{Knecht2013,Hong2016}, the MCP detector is used to detect the impact positions of the recoil $^{6}$Li ions and to provide the stop signal for the ion time-of-flight (TOF) measurement which is triggered by the detection of the $\beta$ particles from $^{6}$He decay. For each event, the ion position and its TOF are needed for reconstructing the initial momentum of the recoil ion and with it the momentum of the outgoing anti-neutrino. The fiducial cut on the MCP image affects the TOF spectral shape and thus the extracted value of the $\beta-\nu$ angular correlation coefficient $a_{\beta\nu}$. It is thus very crucial to calibrate the MCP position response to a high accuracy for such experiments.

To make an MCP detector position-sensitive, a resistive anode \cite{Lampton_1979}, a wedge-and-strip anode (WSA) \cite{siegmund_1983}, or a pair of delay-lines \cite{Roentdek2011} perpendicular to each other are placed behind the MCP to collect the cloud of electrons. A phosphor screen coupled to a CCD camera can also be used, but at the cost of losing high-resolution timing information correlated with position. For the resistive-anode scheme, the electron-collecting position is determined by the ratios of the charge collected at four corners of the resistive anode, while for a WSA anode it is deduced from the ratio of three charges collected on the WSA pattern. For the delay-line scheme, X and Y coordinates are encoded in the time differences between the signals read from the two ends of two perpendicular delay-lines. A resolution of $\approx$100~$\mu$m has been achieved using these anode configurations, and there are also many studies on improving MCP position resolutions for different types of impacting particles \cite{FRASER_1984}, counting rate \cite{Jagutzki_2002} and configurations of MCPs like the number of layers \cite{Siwal_2015}. However, distortions around 200~$\mu$m are inevitable for all position readout schemes \cite{Wiggins_2015,Jagutzki_2002,Lienard2005}. Particularly, distortions are large near the edge of the MCP detector due to fringe electric fields. Therefore a position calibration procedure is needed to achieve a high accuracy. Conventionally, the position response of an MCP detector is calibrated using a mask with holes or other patterns at well-machined positions  \cite{GorelovThesis, Lienard2005, FriedagThesis}. Then the MCP is illuminated with $\alpha$ particles \cite{FriedagThesis}, low-energy ions  \cite{Lienard2005}, low-energy electrons \cite{Wiggins_2015} or ultraviolet photons \cite{Jagutzki_2002} such that the mask pattern is imaged onto the MCP. Usually, a global correction algorithm for the whole MCP is obtained by comparing the imaged hole positions reconstructed by the MCP detector to the actual hole positions on the mask. For example, an accuracy of 240~$\mu$m was achieved for the delay-line position readout scheme \cite{Lienard2005}. If a local correction algorithm is implemented so that the position correction of an event is based on the positions of the holes near this event, the position accuracy can be improved to 120~$\mu$m \cite{FriedagThesis}.

The position accuracies of the conventional calibration schemes described above are limited to about 100~$\mu$m. Moreover, there are two drawbacks of these methods. Firstly, after the calibration is done, the calibration mask needs to be removed, and this involves venting the vacuum chamber and disassembling part of the detector system for the experiment. It is also hard to control the stability of the calibration throughout the experiment when the mask is no longer mounted on the MCP. Secondly, the mask is usually positioned at some distance from the MCP surface, which requires a good understanding of the particle trajectories. This implies a very precise knowledge of the $\alpha$-source position or of the ion flight path which is usually driven by an electric field. 

To overcome these difficulties, we developed a calibration scheme using a 90\% open calibration mask with precisely-shaped orthogonal grids placed directly on top of the MCP surface. In this way, the open area of the mask is large enough so that there is no need to remove the mask after the calibration. The shadow created by the mask is present on the MCP image during the experiment, and thus the position calibration is built into the data. Since there is no gap between the mask and the MCP surface, the requirement of knowing the trajectories of the incoming particles is less strict. In order to achieve a higher position accuracy, we developed an algorithm to determine the positions of the grid lines on the MCP image and then correct the detector position response. In this paper, we will describe the MCP detector system used to develop and test this position calibration method. The position calibration method and its performance will be described in detail. We will also discuss the performance of the MCP detector position response in an experiment using laser-trapped $^{6}$He atoms \cite{Leredde2015}. 

\section{Apparatus and detector operations}
\label{Sec_Apparatus}

\begin{figure}[h!]
\centering
  \includegraphics[width=0.5\linewidth]{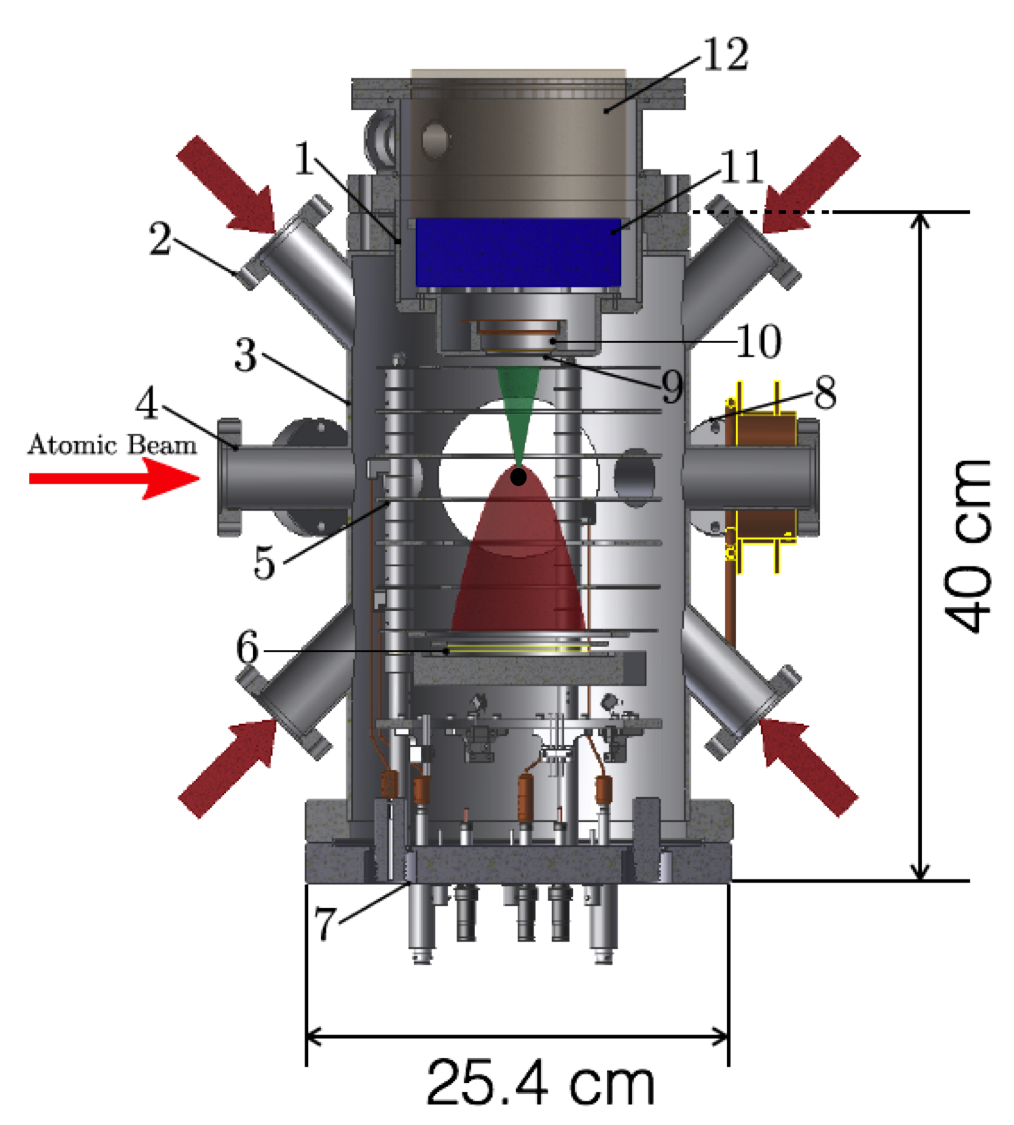}
\caption{Cross-section view of the detector system mounted on the $^{6}$He-trap chamber. 1) re-entrant $\beta$-telescope housing, 2) trapping laser ports, 3) main chamber, 4) $^{6}$He transfer port, 5) electrode assembly, 6) micro-channel plate (MCP) recoil-ion detector, 7) 10~inch custom feedthrough flange for HV and MCP connections, 8) trap monitoring ports, 9) 127~$\mu$m Be foil, 10) multi-wire proportional
chamber (MWPC), 11) plastic scintillator, 12) lightguide to photo-multiplier tube. The black dot at the chamber center is the position of the $^{6}$He trap used in the $\beta-\nu$ correlation experiment and the test with trapped $^{6}$He atoms described in Section~\ref{Sec_Discussion}. The trap is 91~mm above the MCP. The green cone represents the $\beta$-detection solid angle, and the magenta parabola is the envelope of all possible $^{6}$Li ion trajectories. This figure was originally produced in Reference~\cite{DWZThesis}.\label{Fig_DetectorSysSchematic}}
\end{figure}

The experimental setup used to test and calibrate the MCP detector is shown in Figure~\ref{Fig_DetectorSysSchematic}, and this setup is also used in the $^{6}$He $\beta-\nu$ angular correlation measurement \cite{Hong2016}. In the $\beta-\nu$ angular correlation measurement $^{6}$He atoms are laser-trapped at the center of the vacuum chamber shown in Figure~\ref{Fig_DetectorSysSchematic}. A $\beta$-telescope which consists of a multi-wire proportional chamber and a scintillator detector is placed above the trap, and an MCP detector is placed below the trap for detecting recoil ions. Electrodes are installed in between the $\beta$-telescope and the MCP detector to create an electric field of $\approx$1.3~kV/cm and accelerate the recoil ions emitted from the trap towards the MCP detector to $\approx$13~keV. Therefore the ion-collecting solid angle becomes larger and the ions have enough energy to trigger the MCP detector with high and uniform efficiency \cite{Lienard2005}. In this system, calibration sources can be inserted to the trap position via a transportation rod. The MCP tests and calibrations use some or all of this setup in those tasks described in the following sections.

\begin{figure}[h!]
\centering
\includegraphics[width=0.8\textwidth]{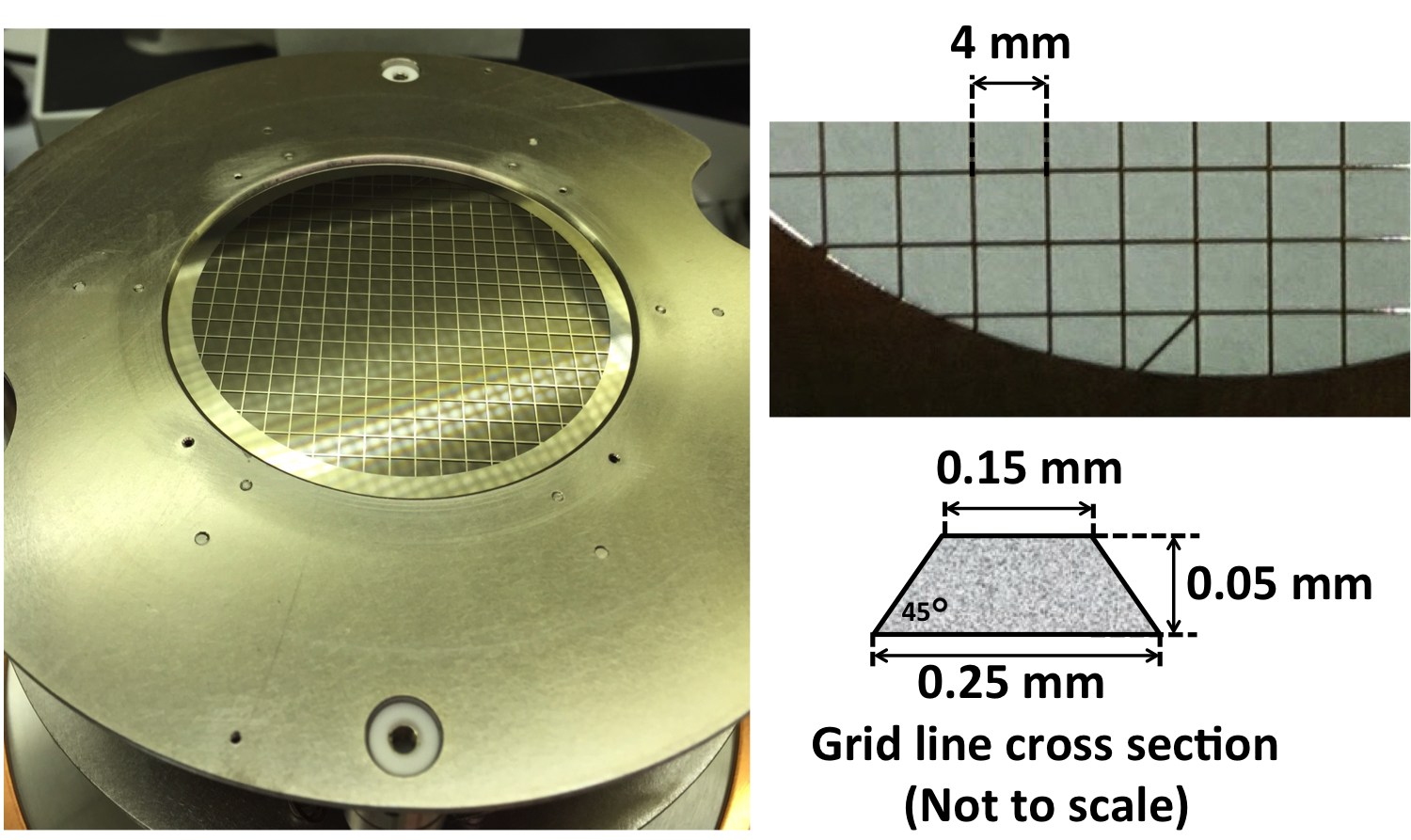}
\caption{Photo of the the micro-channel plate with the calibration mask placed on top of it. The zoomed-in picture of the calibration mask and the grid line cross section are shown on the right. The diagonal line in the bottom square is to mark the orientation of the mask.\label{Fig_MCPMask}}
\end{figure}

\begin{figure}[h!]
\centering
\includegraphics[width=0.9\textwidth]{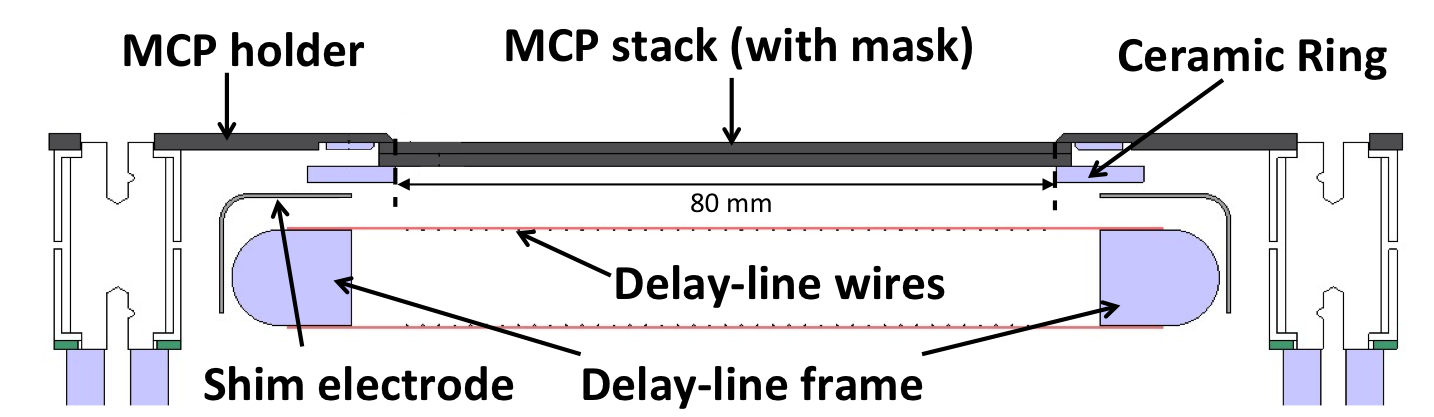}
\caption{Section view of the MCP system schematics. The delay-line wires are wound onto the ceramic delay-line frame, and the wire plane is perpendicular to this cross-section plane. \label{Fig_MCPMaskSchem}}
\end{figure}

The MCP detector system is based on the DLD80 system \cite{Roentdek2011} developed by RoentDek with two MCPs mounted in a chevron configuration. The diameter of each micro-channel is 25~$\mu$m, and the distance between adjacent channels is 35~$\mu$m. The micro-channels are inclined at 8$^{\circ}$ relative to the normal direction of the MCP surface. We have modified the system by mounting the MCP stack to the bottom electrode (the MCP holder) of the recoil-ion spectrometer. The 50~$\mu$m thick nickel calibration mask (Figure~\ref{Fig_MCPMask}) is fixed to the MCP holder and kept flat with four screws, and the MCP stack is clamped to the bottom of the calibration mask by a ceramic ring as shown in Figure~\ref{Fig_MCPMaskSchem}. The MCP holder, the calibration mask, and the top surface of the MCP stack are in electrical contact with each other. The shim electrode and the delay-line anodes are adjusted so that the MCP stack is at the position recommended for the DLD80 system by RoentDek. The calibration mask is produced through {\em electroforming} \cite{VecoFrance}, and nickel is chosen for sake of its stiffness. The grid lines of the calibration mask are separated by 4~mm, and the width of each grid line is 250~$\mu$m. The accuracy of the positions and widths of grid lines is 2~$\mu$m according to the manufacturer, and this was confirmed using a {\em SmartScope ZIP Lite 250} mechanical inspection system. The grid lines are approximately parallel to the directions of the X and Y delay-line wires. The edges of the grid lines are formed with a $45^{\circ}$ inclination relative to the mask surface as shown in Figure~\ref{Fig_MCPMask}. Therefore, incoming particles with impact angles smaller than $45^{\circ}$ are not affected by the thickness of the calibration mask. The inner diameter of the mask is 75~mm which defines the detection area for ions and $\alpha$ particles. The overall dark rate of our MCP is about 0.01~count$\cdot$s$^{-1}\cdot$mm$^{-2}$. We observed that the dark rate increased by $\approx$15\% on average across the whole usable surface of the MCP (up to 65\% in some regions) after the calibration mask was installed. This increase is possibly caused by charges released from sharp points on the mask or ionizations of residual gas. 

The front side of the MCP assembly is biased through the MCP holder, while the back side is biased through a conductive layer on the ceramic ring. The bias voltages applied to the front and back side of the MCP, the shim electrode and the delay-line anode wires are listed in Table~\ref{Tab_MCPVoltages}. Between the front and back of the MCP, a bias voltage of $\approx$2400~V is applied, and the overall gain of the two MCPs is $\sim$10$^{7}$. The avalanche of electrons leaving the MCPs induces a fast positive signal on the back side of the MCPs and is collected by the {\em signal} wires of the X and Y delay lines. The {\em reference} wires wound next to the {\em signal} wires are biased with slightly different voltages and are used to suppress the electromagnetic noise picked up in the vacuum chamber \cite{Roentdek2011}. Five signals are read out: one from the MCP back side, two from the ends of the X delay line, and two from the ends of the Y delay line. The MCP back-side signal is inverted and then amplified by an Ortec VT120C fast amplifier. The four anode signals are amplified by an Ortec FTA 420C four-channel fast amplifier. The gain of all these amplifiers is 20. We use the {\em Fast Acquisition SysTem for nuclEar Research} (FASTER) \cite{FASTER} as our data acquisition (DAQ) system. Each one of the five amplified signals is digitized by a 500~MHz flash analog-to-digital converter (flash ADC). All flash ADCs are synchronized, and the pulse filtering, timing and charge integrating are performed in hardware using modular algorithms loaded onto the field-programmable gate arrays (FPGAs) of the DAQ system. The pulse filtering includes a low-pass filter with a time constant of 13~ns and a baseline restoration (BLR) treatment which applies a 160~kHz low-pass filter to the input  waveform and subtracts this filtered waveform. The BLR is vetoed when the input waveform is above a 15~mV threshold and during a 1~$\mu$s gate after the input returns below the threshold. The timing of a pulse, which determines the MCP detector position readout, is extracted using the constant-fraction discrimination algorithm. A timing resolution of $\approx$50~ps was achieved for the MCP detector and the acquisition system used. Charge integrating within a time window of 32~ns carries additional information on the pulse amplitude for each signal. 

\begin{table}[h]
\centering
\label{my-label}
\begin{tabular}{cc}
\hline
Electrode      & Voltage (V) \\ \hline
MCP Top      & -2000       \\
MCP Bottom       & 371       \\
Shim Electrode & 400        \\
Reference Wire & 762       \\
Signal Wire    &   816       \\ \hline
\end{tabular}
\caption{Voltages of the ion detector system.\label{Tab_MCPVoltages}}
\end{table}

The position where the incoming particle hits the MCP detector is determined by the time difference between the two signals from each end of the same delay-line anode. Two sets of delay-lines are wound at 1~mm pitch, and their directions are orthogonal with respect to each other. It takes $\sim$2~ns for a signal to travel through one turn of the delay-line, so the MCP position readout (X,Y) can be expressed as 
\begin{align}
\label{Eq_MCPPosRec}
X&=\frac{TX_{1}-TX_{2}}{2\text{ ns/mm}}\times k_{X}+c_{X},\nonumber\\
Y&=\frac{TY_{1}-TY_{2}}{2\text{ ns/mm}}\times k_{Y}+c_{Y},
\end{align}
where $TX_{1}$, $TX_{2}$, $TY_{1}$, $TY_{2}$ are the timings of the four delay-line anode signals in nanoseconds. The ns-to-mm conversion factor is a priori not known with high accuracy, and timing offsets are easily introduced in the pulse propagation and amplification. Therefore, the scaling factors $k_{X}$, $k_{Y}$ and shifts $c_{X}$, $c_{Y}$ are introduced in the position reconstruction formula for the basic stretching and shifting. These parameters need to be determined through calibration. Although this basic correction scheme cannot restore rotations and non-linear distortions of the MCP image, it makes the high-precision calibration algorithm easier to implement.

\section{Position calibration}
\label{Sec_PosCal}

To calibrate the MCP position response, we use high-intensity $\alpha$ sources or low-energy ion-sources to illuminate the MCP, and collect enough events (more than $10^{7}$) so that the mask grid line edges on the MCP image can be resolved with a precision better than 10~$\mu$m. Then the positions of the grid line crossings and several other points along the grid line in between the crossings are determined. A correction function is constructed for each square region enclosed by grid lines in order to shift the positions of the points determined above as close as possible to their physical positions. These functions are then used to correct the MCP position response on an event-by-event basis for other runs. In this section, details of this calibration procedure are described. The accuracy and resolution of the calibrated MCP detector position response are determined. The example used for illustrating the calibration procedure is based on calibration runs using $\alpha$ particles from a $^{241}$Am source placed at the trap position shown in Figure~\ref{Fig_DetectorSysSchematic}. Calibrations using $^{6}$Li ions from decays of non-trapped $^{6}$He atoms are described at the end of this section.

\subsection{Event selection and pre-calibration}
\label{SubSec_PreCal}

\begin{figure}[h!]
\centering
\subfloat[]{\includegraphics[width=0.49\textwidth]{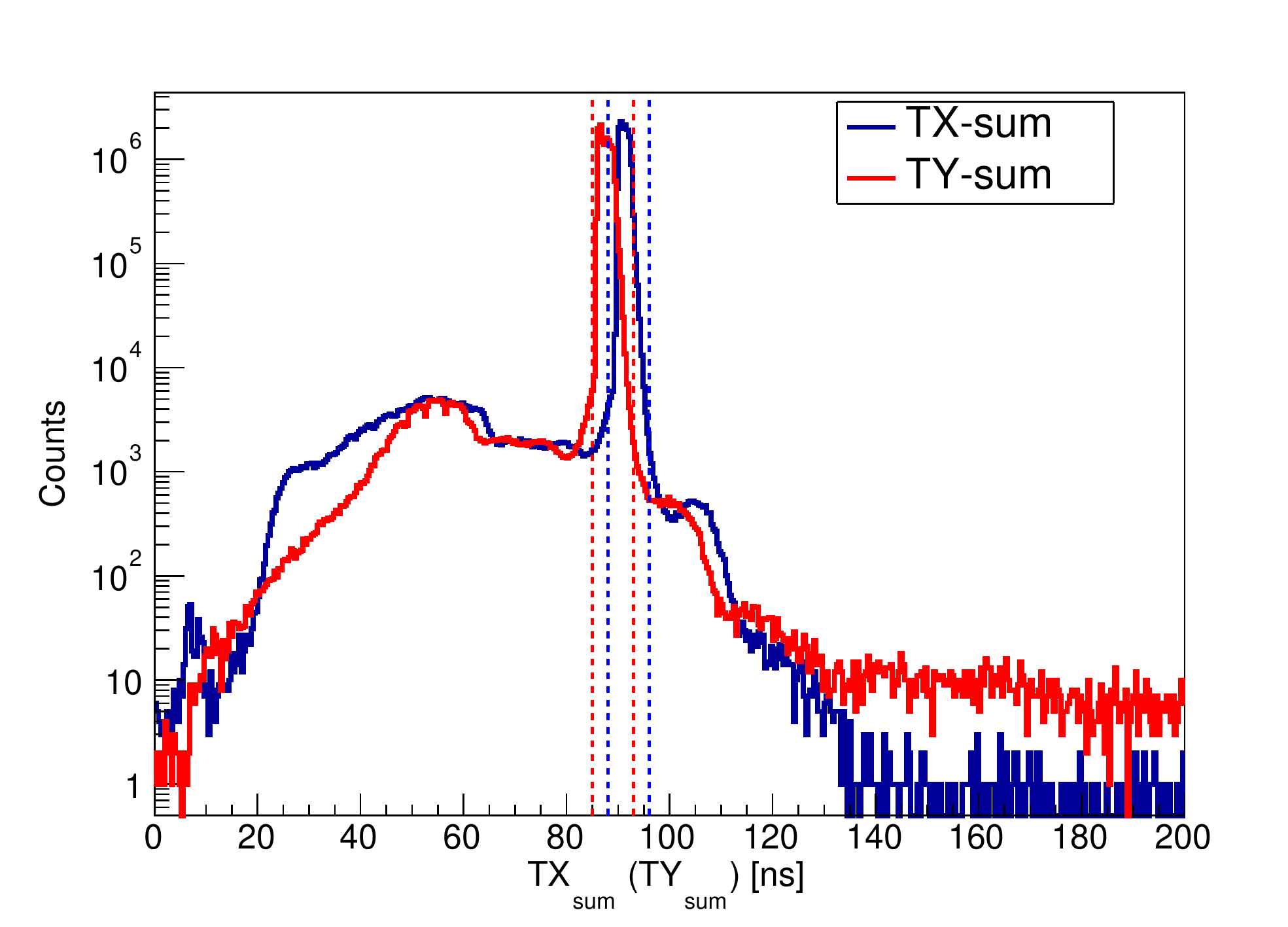}\label{Fig_TimeSumMCP}}~~
\subfloat[]{\includegraphics[width=0.49\textwidth]{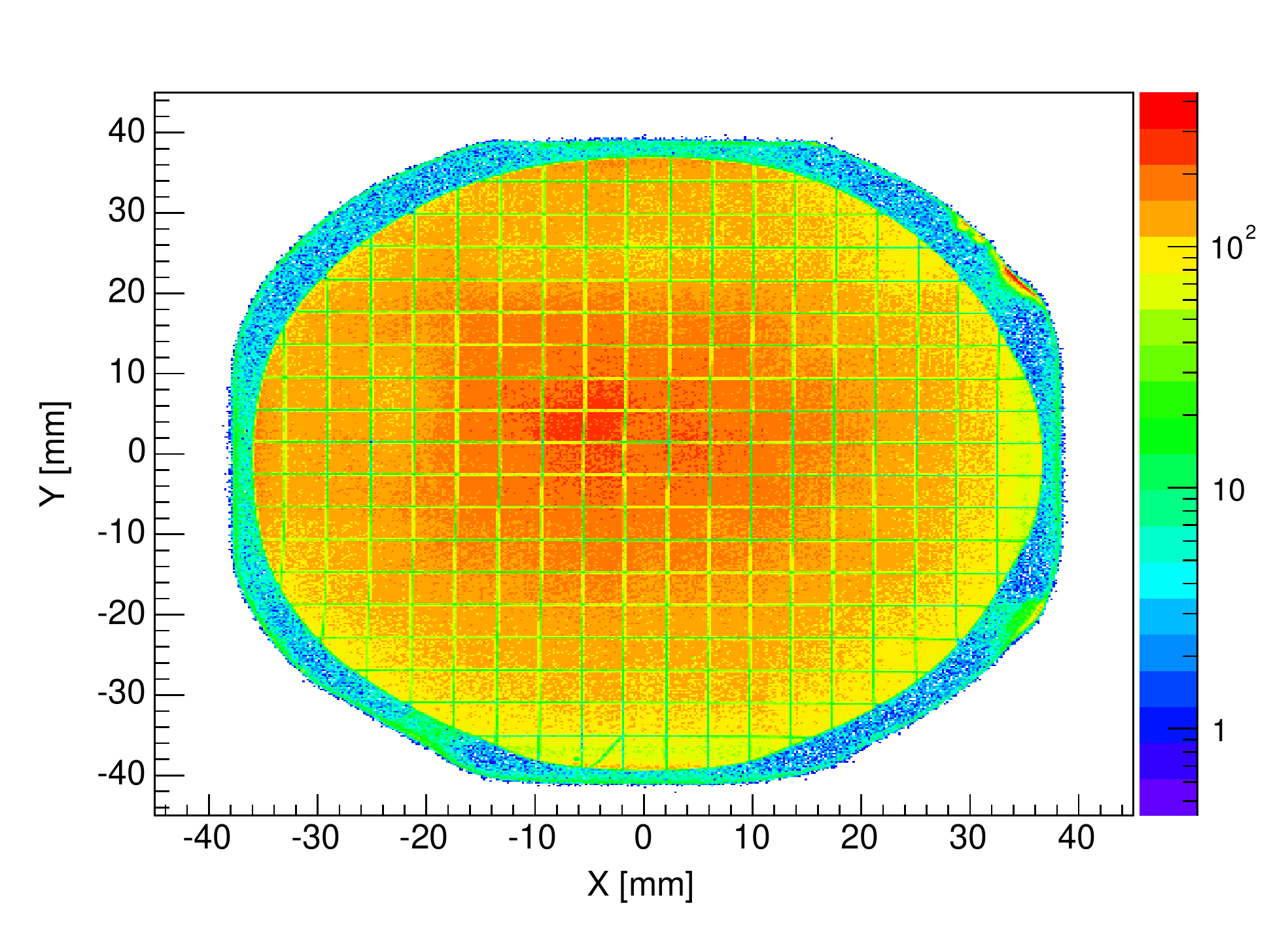}\label{Fig_MCPPosUncal}}
\caption{(a) $TX_{sum}$ (Blue) and $TY_{sum}$ (Red) spectra with cut region indicated. (b) MCP image with the cut on $TX_{sum}$ and  $TY_{sum}$. This image is constructed without any corrections described in Subsection~\ref{SubSec_PosCorrection}.}
\end{figure}

Before conducting the calibration, one condition on the timing of the signals is imposed for event selection in order to rule out false triggers by electronic noise and pile-up events. Because the total length of the delay-line is fixed, the sum of the propagation times from the charge impact position to the two ends of the delay-line is fixed, independent of where the event happens. Therefore, the timing sums
\begin{align}
TX_{sum}&=(TX_{1}-T_{MCP})+(TX_{2}-T_{MCP}),\nonumber\\
TY_{sum}&=(TY_{1}-T_{MCP})+(TY_{2}-T_{MCP}),
\label{Eq_TimeSumMCP}
\end{align}
are both expected to be constants, where $T_{MCP}$ is the timing of the MCP back-side signal. The spectra of $TX_{sum}$ and $TY_{sum}$ for the calibration run are shown in Figure~\ref{Fig_TimeSumMCP}. The sharp peaks near 90~ns in these spectra agree with the expectation described above. Events with $TX_{sum}$ or $TY_{sum}$ away from these peaks must have false triggers in some channels, and thus are ruled out by applying a $\pm$4~ns acceptance window around the peaks. Events cut out by imposing this condition are $\sim$5\% of the whole data set, and this condition is applied for all spectra shown in the rest of this paper.

The 2D spectrum of the $\alpha$-hitting positions (the MCP image) is shown in Figure~\ref{Fig_MCPPosUncal}. All grid lines, their crossings, and the inner rim of the mask are visible. The $k$ and $c$ parameters in Equation~\ref{Eq_MCPPosRec} are chosen such that the middle square region is centered at the origin of the coordinate system and its edges are all 4~mm. Even without any further corrections, the quality of this position reconstruction is good enough for many applications \cite{Gorelov.142501,Flechard_2012}: the grid line crossings in Figure~\ref{Fig_MCPPosUncal} deviate by $\approx$250~$\mu$m from their physical positions on average and up to 1.1~mm near the edge, as will be described in Section~\ref{SubSec_GridFinding}. Our high-accuracy calibration is based on this image.

\subsection{Grid line finding algorithm}
\label{SubSec_GridFinding}

\begin{figure}[h!]
\centering
\subfloat[]{\includegraphics[width=0.49\textwidth]{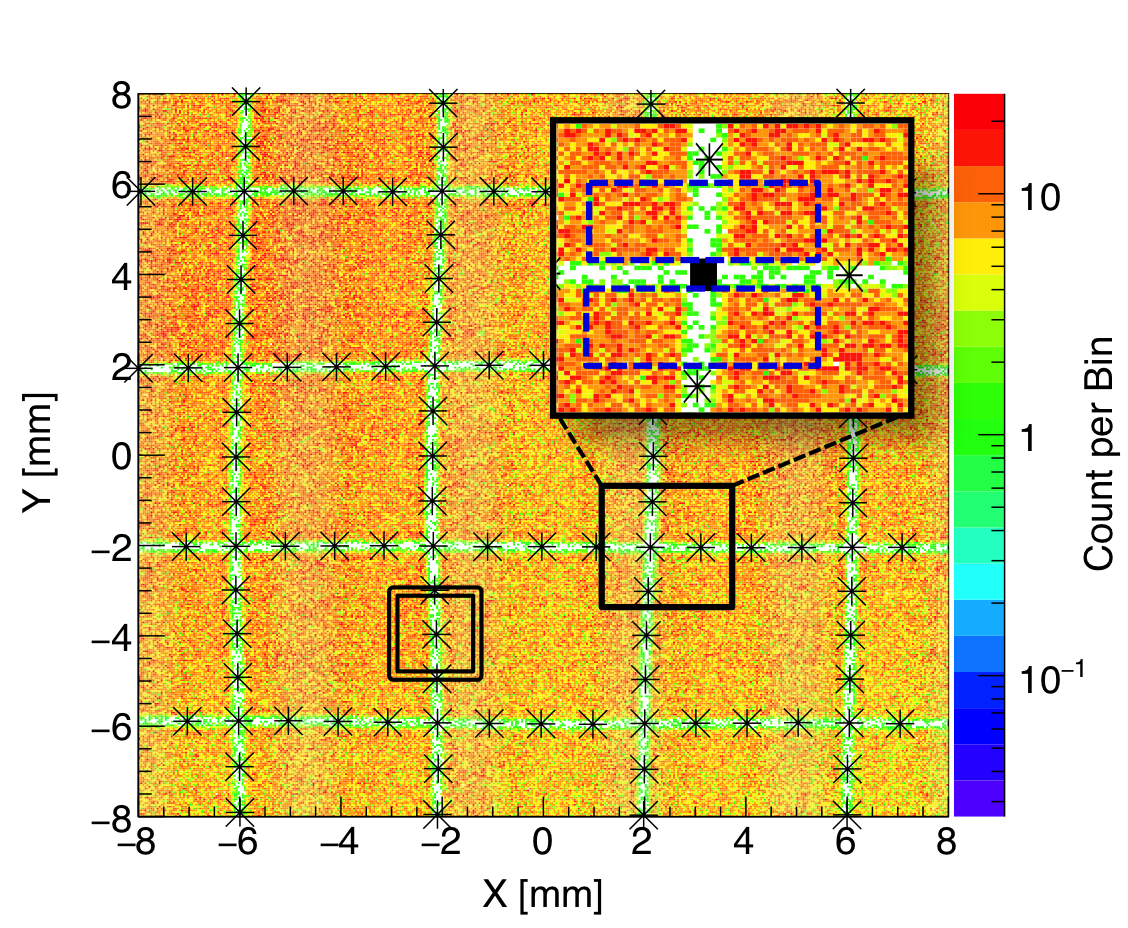}\label{Fig_MCPGridOn}}
\subfloat[]{\includegraphics[width=0.49\textwidth]{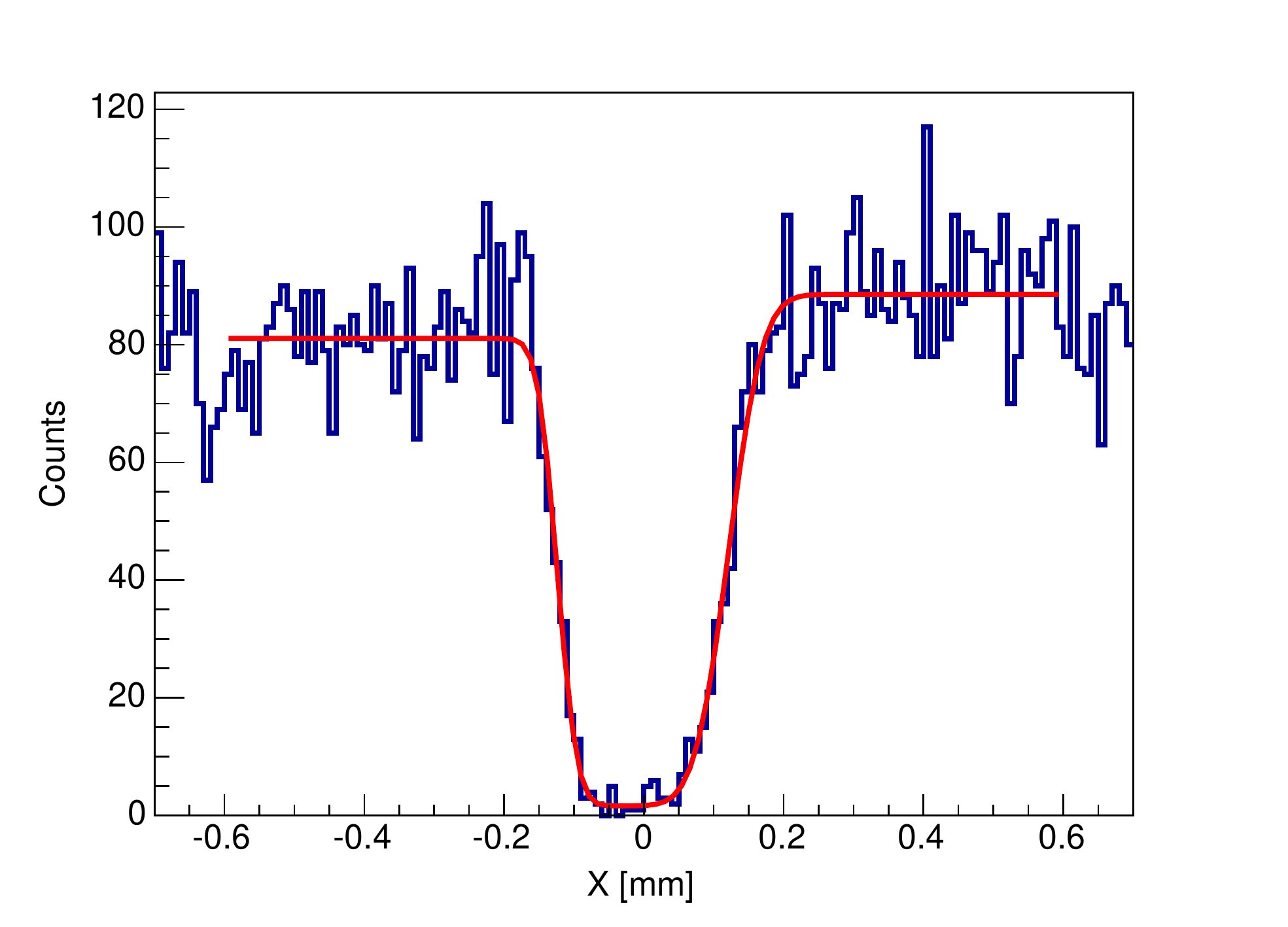}\label{Fig_LocalProjection}}
\caption{(a) Calibration points (marked as ``$\ast$'') determined along the grid line shadow in the MCP image. The rectangular region at (2, $-$2) is magnified in the insert, and the blue dashed rectangles in the insert are the projection regions for determining the X coordinate of the grid line crossing point shown as the black square dot. (b) Projection of the region (black rectangle with double-lined edges in (a)) around ($-$2, $-$4) onto the X axis. The X coordinate in this histogram is relative to $X=-2$. The red line is the fit function defined in Equation~\ref{Eq_MCPGridResponse}. Projections of the blue dashed rectangles in the insert are similar to this histogram.}
\end{figure}

In order to correct the MCP detector position readout, one needs to determine the grid line positions on the MCP image with high accuracy. First of all, we determine the positions of grid line crossings. After the scaling and shifting described above, the grid line crossings are approximately at their physical positions within up to 2~mm. We make initial guesses of the grid line crossings, and then search for the local minima of the horizontal and vertical projections of the MCP image in a 1.6~mm by 1.6~mm square region centered at the initially-guessed positions of each grid line crossing. The low-accuracy determinations of the grid line crossing coordinates are assigned to the positions of these local minima. To achieve a higher-accuracy determination, for each grid line crossing we project onto the horizontal axis the 1.6~mm by 0.8~mm rectangular regions (the blue dashed rectangles in the insert of Figure~\ref{Fig_MCPGridOn}) above and below its low-accuracy position determined previously. These projections have similar shapes to the spectrum shown in Figure~\ref{Fig_LocalProjection}. Ideally, the projection spectrum should be a step function at either edge of a grid line. Due to the finite resolution of the MCP, the step function becomes an error function. Therefore, we fitted these projections with a function $F(x)$ defined as
\begin{align}
F(x)=N_{1}\times \left(1-\text{Erf}\left(\frac{x-\mu_{1}}{\sqrt{2}\sigma_{1}}\right)\right)+N_{2}\times \left(1+\text{Erf}\left(\frac{x-\mu_{2}}{\sqrt{2}\sigma_{2}}\right)\right)+B,
\label{Eq_MCPGridResponse}
\end{align}
where $N_{1}$ and $N_{2}$ are normalization factors, $B$ is a constant to account for the dark rate and events triggered by particles going through the mask such as cosmic rays, $\mu_{1}$ and $\mu_{2}$ are the positions of the falling edge and the rising edge respectively, and $\sigma_{1}$ and $\sigma_{2}$ are the Gaussian smearing widths of the edges to account for the position resolution. The error function Erf($x$) is defined as:
\begin{align}
\text{Erf}(x)&=\frac{2}{\sqrt{\pi}}\int_{0}^{x}\exp (-t^{2})dt.
\label{Eq_ErrorFunctions}
\end{align}
The fitted function is shown in red in Figure~\ref{Fig_LocalProjection}. The statistical uncertainties of the $\mu$ parameters are all below 6~$\mu$m. The high-accuracy determination of the X coordinate of a grid line crossing ($X_{cross}$) is the average of the fitted $\mu$ values for the projection histograms:
\begin{align}
X_{cross}=\frac{\mu_{1}^{above}+\mu_{2}^{above}+\mu_{1}^{below}+\mu_{2}^{below}}{4}.
\end{align}
The high-accuracy determination of the Y coordinate of a grid line crossing ($Y_{cross}$) is determined in the same manner using the vertical projections of the regions on the left and right sides of the grid line crossing position. Secondly, after having determined the coordinates of all grid line crossings with high accuracy we select three more points on each segment of the grid lines as shown in Figure~\ref{Fig_MCPGridOn}. For each horizontal segment, X coordinates of the three intermediate points are set uniformly in between the grid line crossings at each end of the segment, and the initial guesses of their Y coordinates are assigned to the average of the Y coordinates of the segment ends. The high-accuracy Y coordinates of these three points are determined by fitting the vertical projection of the 1.6~mm by 1.6~mm region around each point and averaging the $\mu$ parameters. The positions of the three intermediate points along vertical segments are determined in the same way. Thus, positions of all {\em calibration points} which include the grid line crossings and the three intermediate points on each grid line segment are determined with high accuracy. 

\begin{figure}[h!]
\centering
\includegraphics[width=0.5\textwidth]{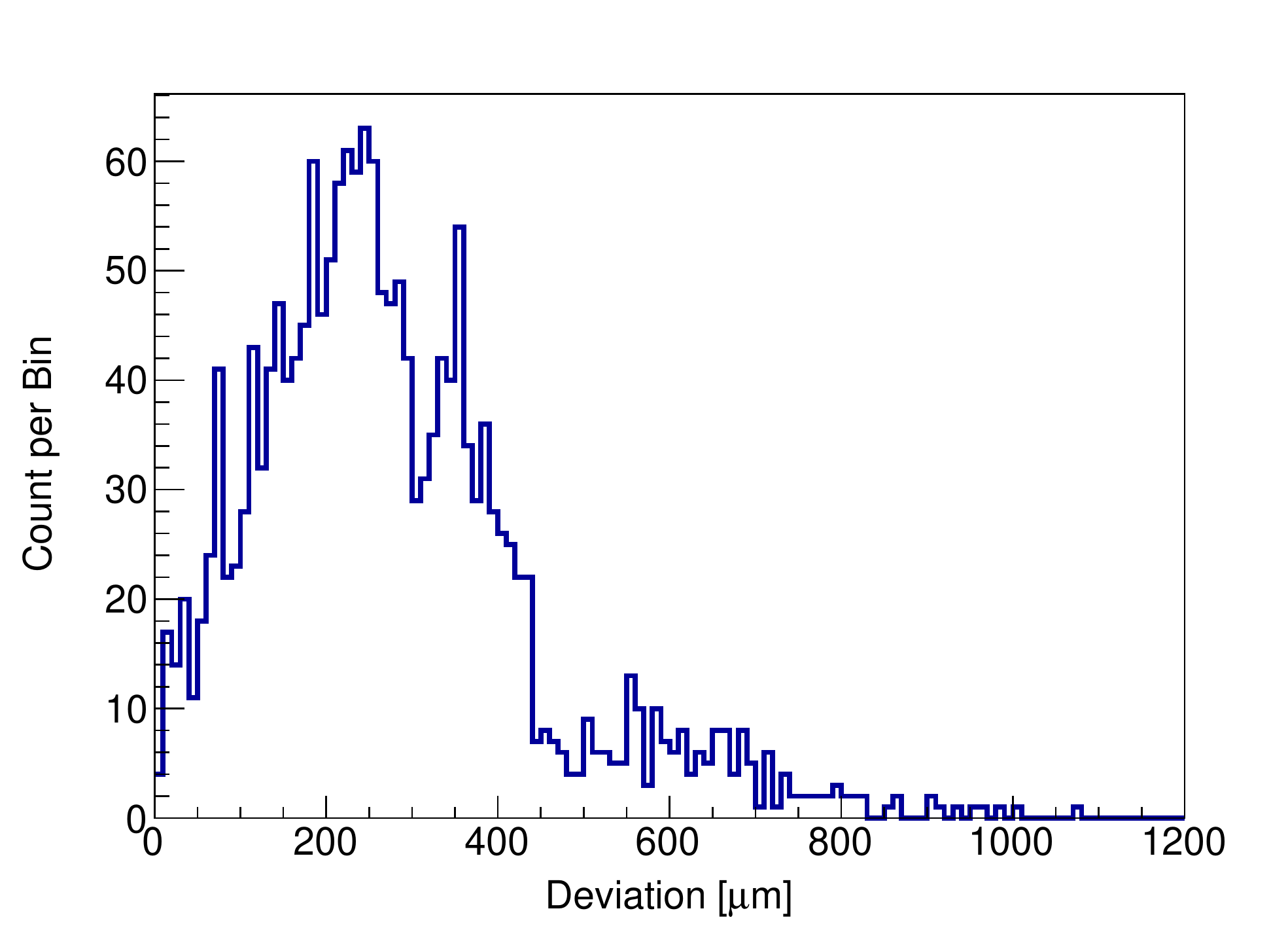}
\caption{Distribution of the absolute deviations of calibration points.\label{Fig_MCPGridDev}}
\end{figure}

To estimate the distortions of the MCP image, we calculate the deviations between the coordinates of the calibration points on the MCP image and their physical positions. The distribution of the absolute deviations is plotted in Figure~\ref{Fig_MCPGridDev}. The mean absolute deviation is 278~$\mu$m and the maximum deviation is 1.1~mm. Observing the whole MCP image, we found an overall rotation of $\approx3\times10^{-3}$~rad, undulating of grid lines, and inward compression near the edge of the MCP. In general, the farther the position is from the center, the larger deviation it exhibits. 

\subsection{Event-by-event position correction}
\label{SubSec_PosCorrection}

For each square region, we use second-order two-variable polynomials to transform the high-accuracy positions ($X$,$Y$) of the calibration points on the square sides to their physical positions ($X_{Phys}$,$Y_{Phys}$) as closely as possible, in order to correct the MCP position readout for each event. The correction functions are
\begin{align}
X_{C}&=p_{0}X^{2}+p_{1}XY+p_{2}Y^{2}+p_{3}X+p_{4}Y+p_{5},\nonumber\\
Y_{C}&=q_{0}X^{2}+q_{1}XY+q_{2}Y^{2}+q_{3}X+q_{4}Y+q_{5}.
\label{Eq_SquareCorrection}
\end{align}
In Equation~\ref{Eq_SquareCorrection}, ($X_{C}$,$Y_{C}$) is the corrected position of a calibration point with non-corrected position ($X$,$Y$), and $p_{i}$ and $q_{i}$ ($i=0\to5$) are parameters determined by a fit using all the calibration points on this square region and setting ($X_{C}$,$Y_{C}$) to their physical positions. The fitted values of these parameters for all complete square regions are saved to a file for the event-by-event position correction of other runs. The regions near the outer edge which are not complete squares are ruled out from the analysis. This only reduces the usable area of the MCP by $\sim$10\%, but avoids the use of complex correction algorithms for regions with non-square shapes. 

\begin{figure}[h!]
\centering
\includegraphics[width=0.5\textwidth]{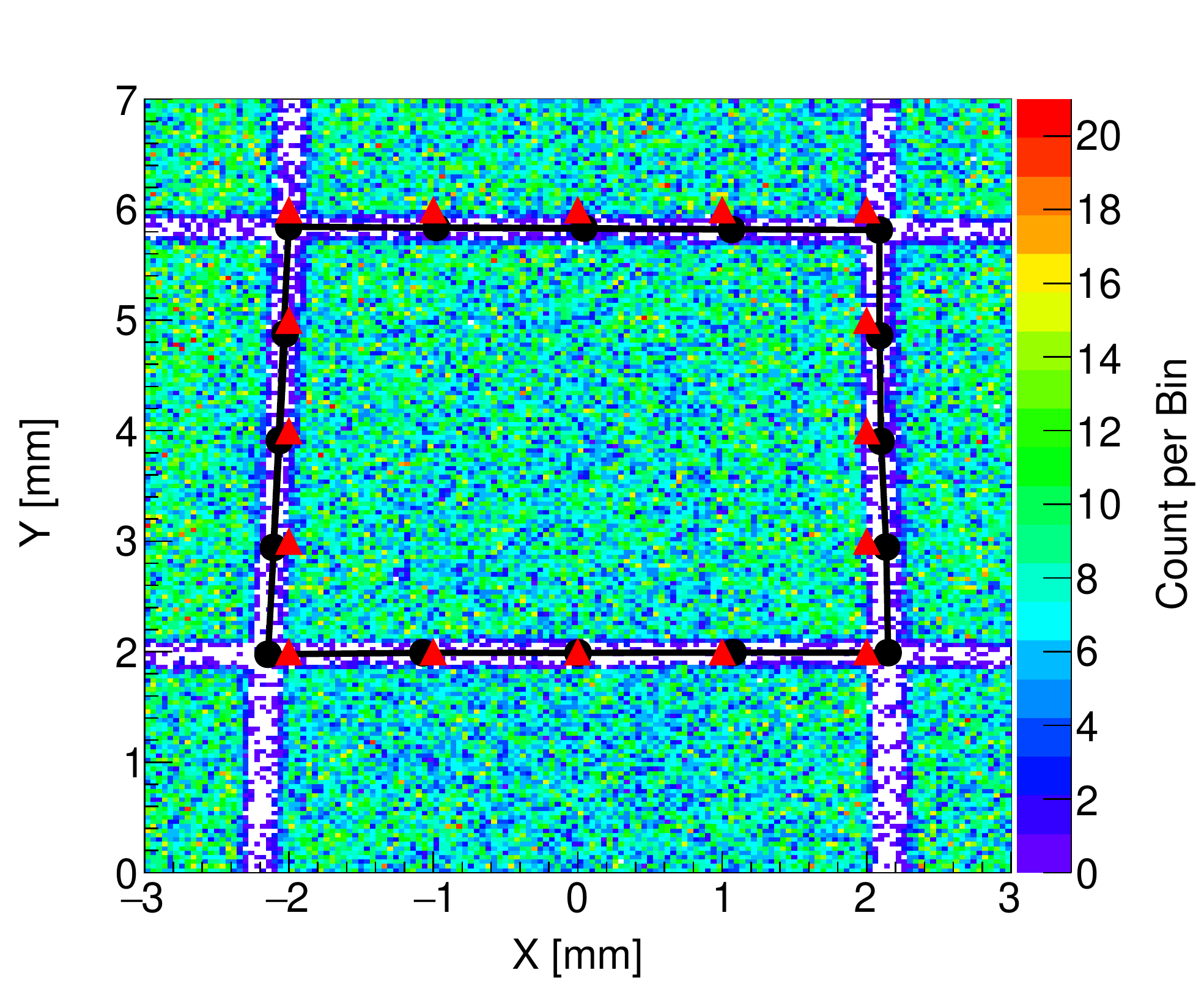}
\caption{The square region centered at (0.0, 4.0) with the calibration points at the corners and on the sides (labeled in black circles and joined with black lines). The physical positions of these points are labeled in red triangles. The black loop indicates the boundary of this region. \label{Fig_LocalSquareCorrection}}
\end{figure}

For the event-by-event position correction, the non-corrected position of one event is first reconstructed using Equation~\ref{Eq_MCPPosRec} with parameters $k$ and $c$ determined as described above. To implement the position correction function, the square region to which this event belongs needs to be determined. The judgement of whether the event belongs to one square region is made by testing whether its non-corrected position is inside the loop created by linking the calibration points on the sides of this square region (from the non-calibrated MCP image of the calibration run) as shown in Figure~\ref{Fig_LocalSquareCorrection}. In this way, the square region that each event belongs to is unambiguously determined. Then the position of this event is corrected using Equation~\ref{Eq_SquareCorrection} with the parameter values associated with its corresponding square region. The quality of this position correction method is discussed in the next subsection.

\subsection{Accuracy and resolution of the calibrated MCP position response}
\label{SubSec_AccuracyResolution}

\begin{figure}[h!]
\centering
\includegraphics[width=0.5\textwidth]{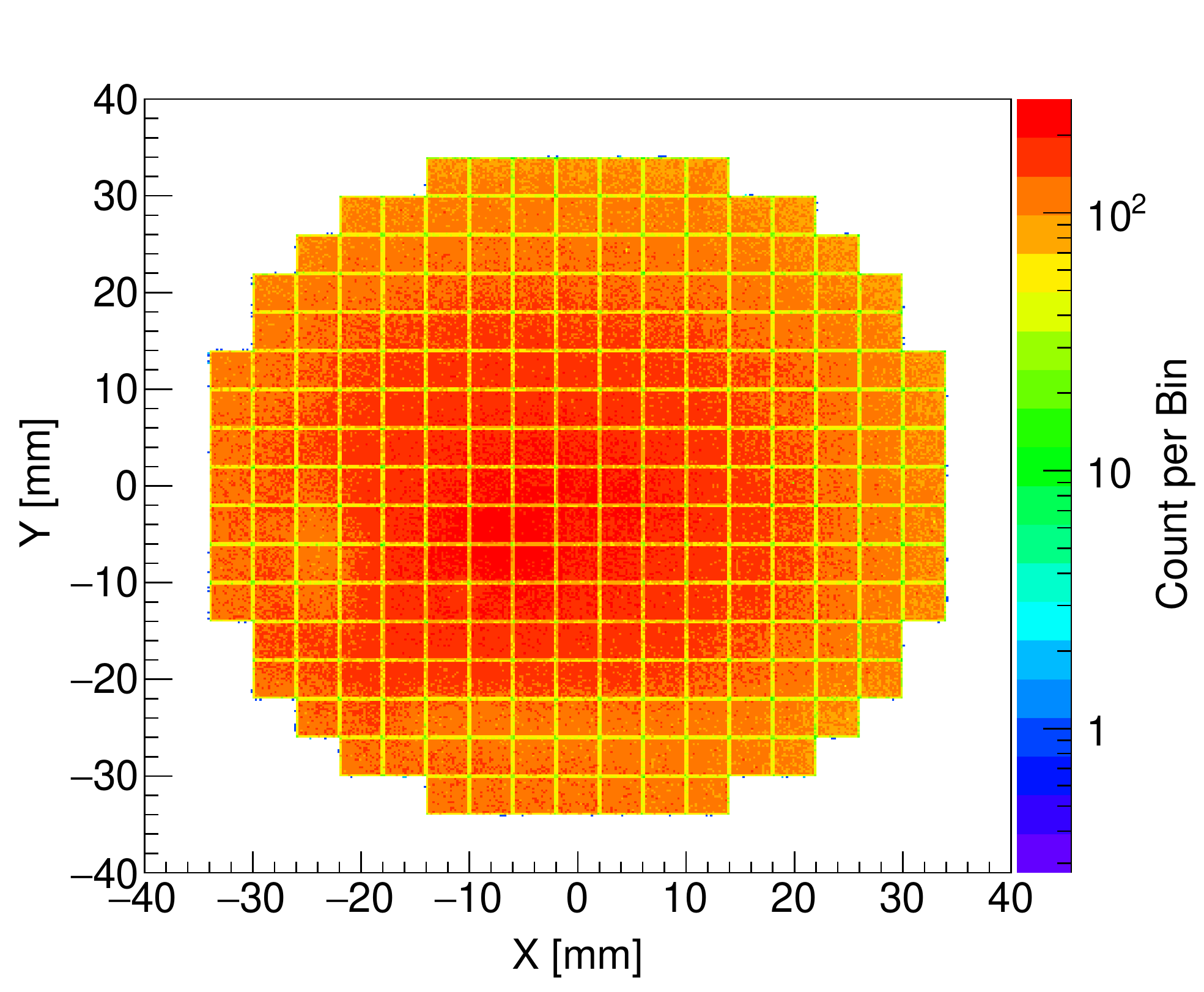}
\caption{MCP position spectrum after the event-by-event position correction. \label{Fig_MCPPosCal}}
\end{figure}

To test the performance of this position calibration method, we processed another data set and corrected the MCP position response using the parameters ($k$'s, $c$'s, $p$'s and $q$'s in Equation~\ref{Eq_MCPPosRec} and Equation~\ref{Eq_SquareCorrection}) determined by the calibration run. This data set was taken under the same conditions as but statistically independent from the calibration run. Therefore, it is appropriate to use this data set to extract the MCP position resolution and the accuracy of this calibration scheme. The MCP image after position corrections is shown in Figure~\ref{Fig_MCPPosCal}. The events falling into the non-square regions near the mask rim are already discarded. 

\begin{figure}[h!]
\centering
\subfloat[]{\includegraphics[width=0.49\textwidth]{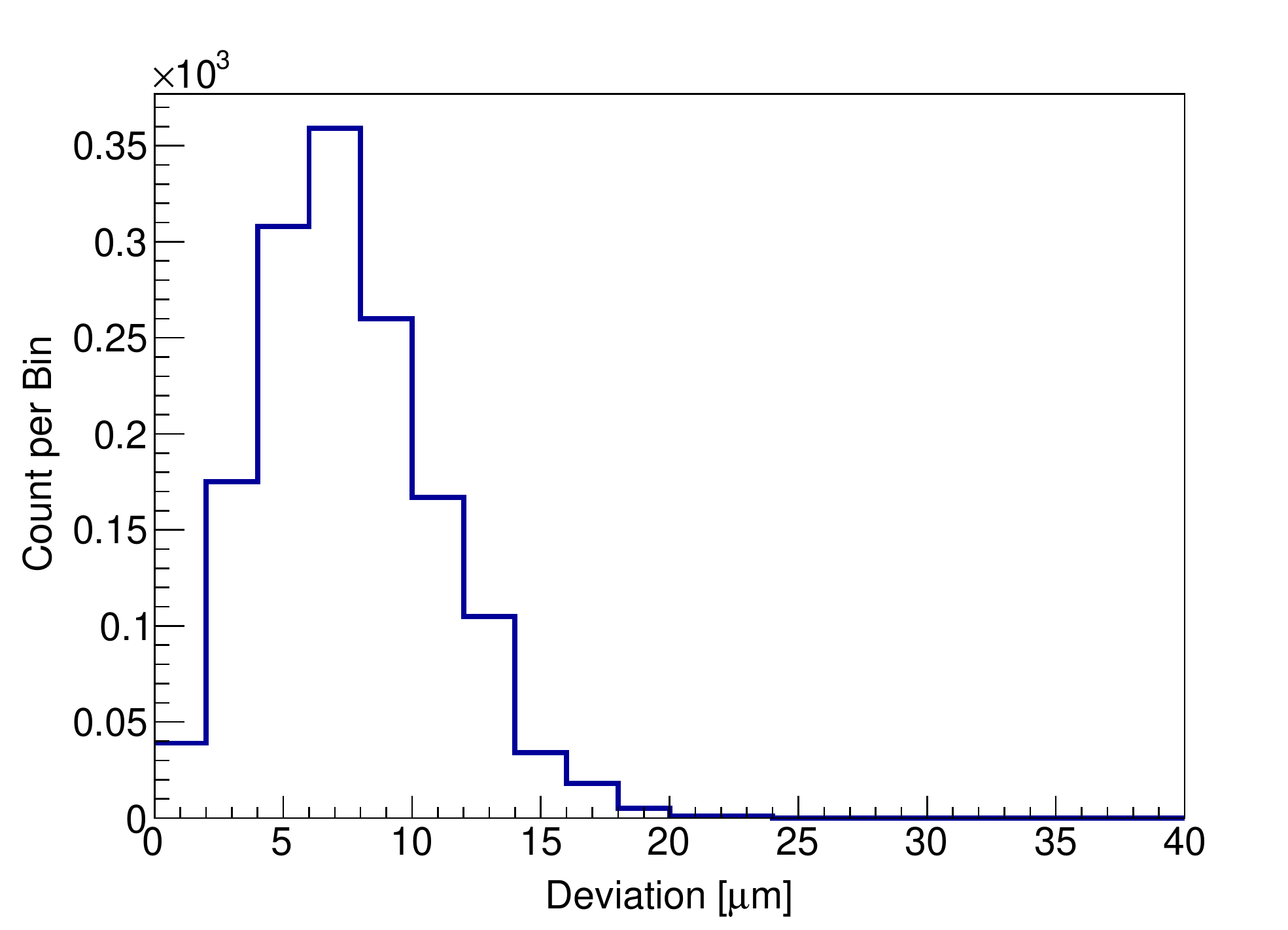}~~\label{Fig_MCPCorrectedDev}}
\subfloat[]{\includegraphics[width=0.49\textwidth]{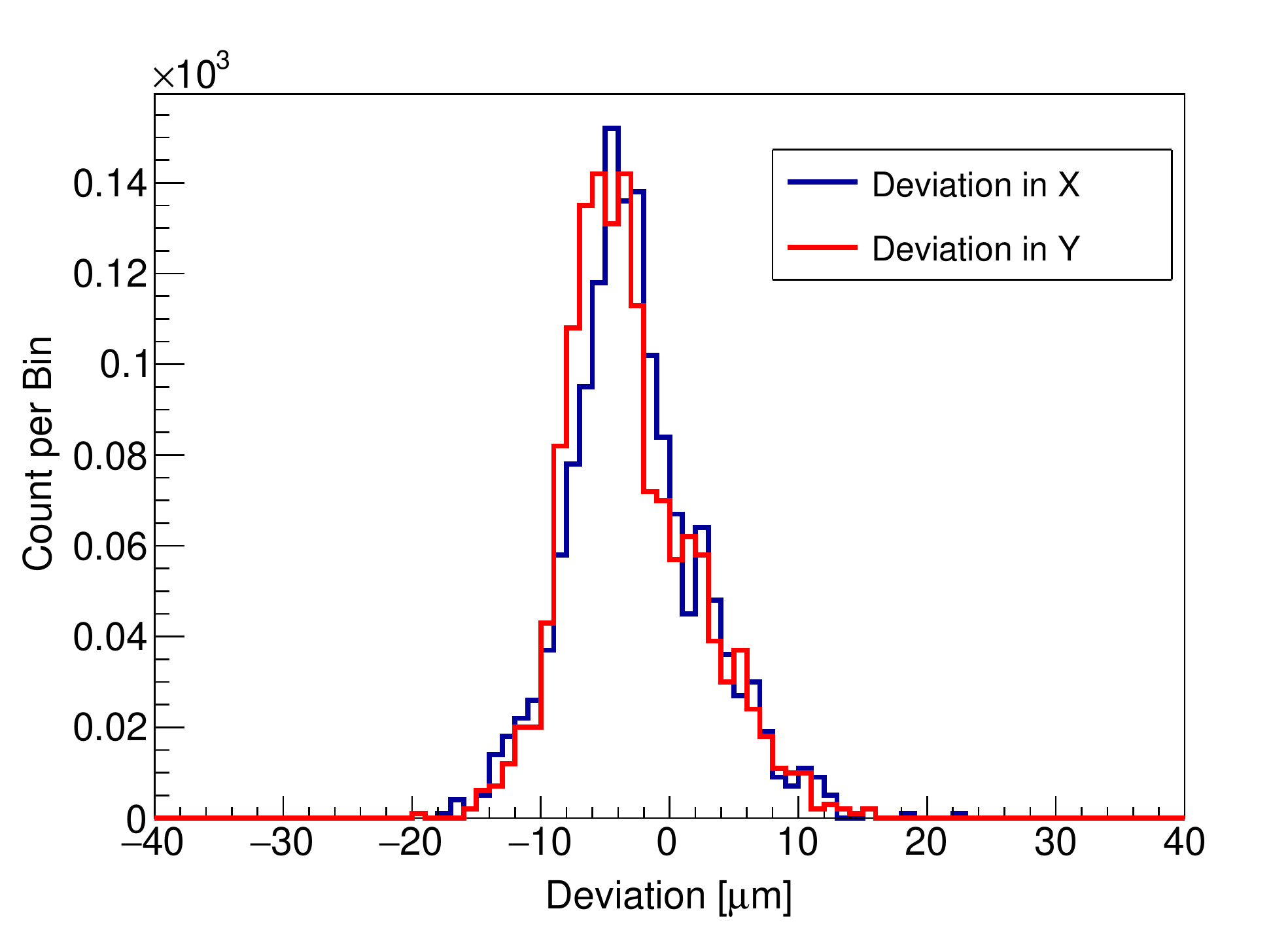}\label{Fig_MCPGridCrossingOverlay}}
\caption{(a) Distribution of the absolute deviations of all grid line crossings and 3 intermediate points along each grid line segment in between the crossings on the corrected MCP image. (b) Distribution of the deviations in the X and Y directions. The mean deviations in both directions are $\approx-$3~$\mu$m.\label{Fig_MCPCorrectedAll}}
\end{figure}

Using the same algorithm as described above, we determined the coordinates of grid line crossings and of three intermediate points along each grid line segment in between the crossings based on the corrected MCP image in Figure~\ref{Fig_MCPPosCal}. Then the deviations of these coordinates from their physical values were calculated. The distributions of the absolute deviations and the deviations in the X and Y directions are shown in Figure~\ref{Fig_MCPCorrectedAll}. The average absolute deviation amounts to 8~$\mu$m, much smaller than it was prior to the correction. The root-mean-square (RMS) of this distribution is 3~$\mu$m, and in this test the maximum deviation is 20~$\mu$m.

\begin{figure}[h!]
\centering
\subfloat[Horizontal grid line segments]{\includegraphics[width=0.49\textwidth]{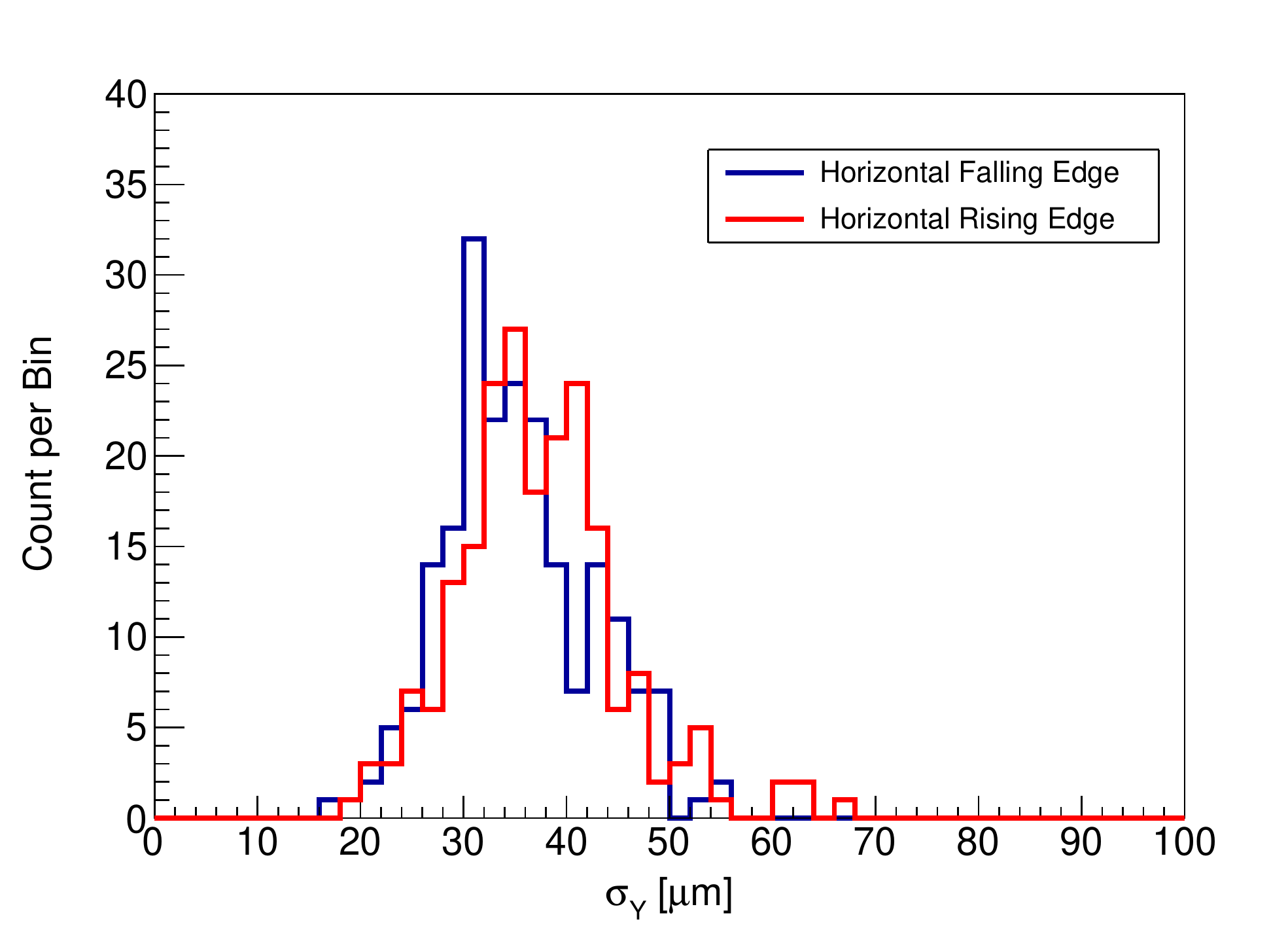}}~~
\subfloat[Vertical grid line segments]{\includegraphics[width=0.49\textwidth]{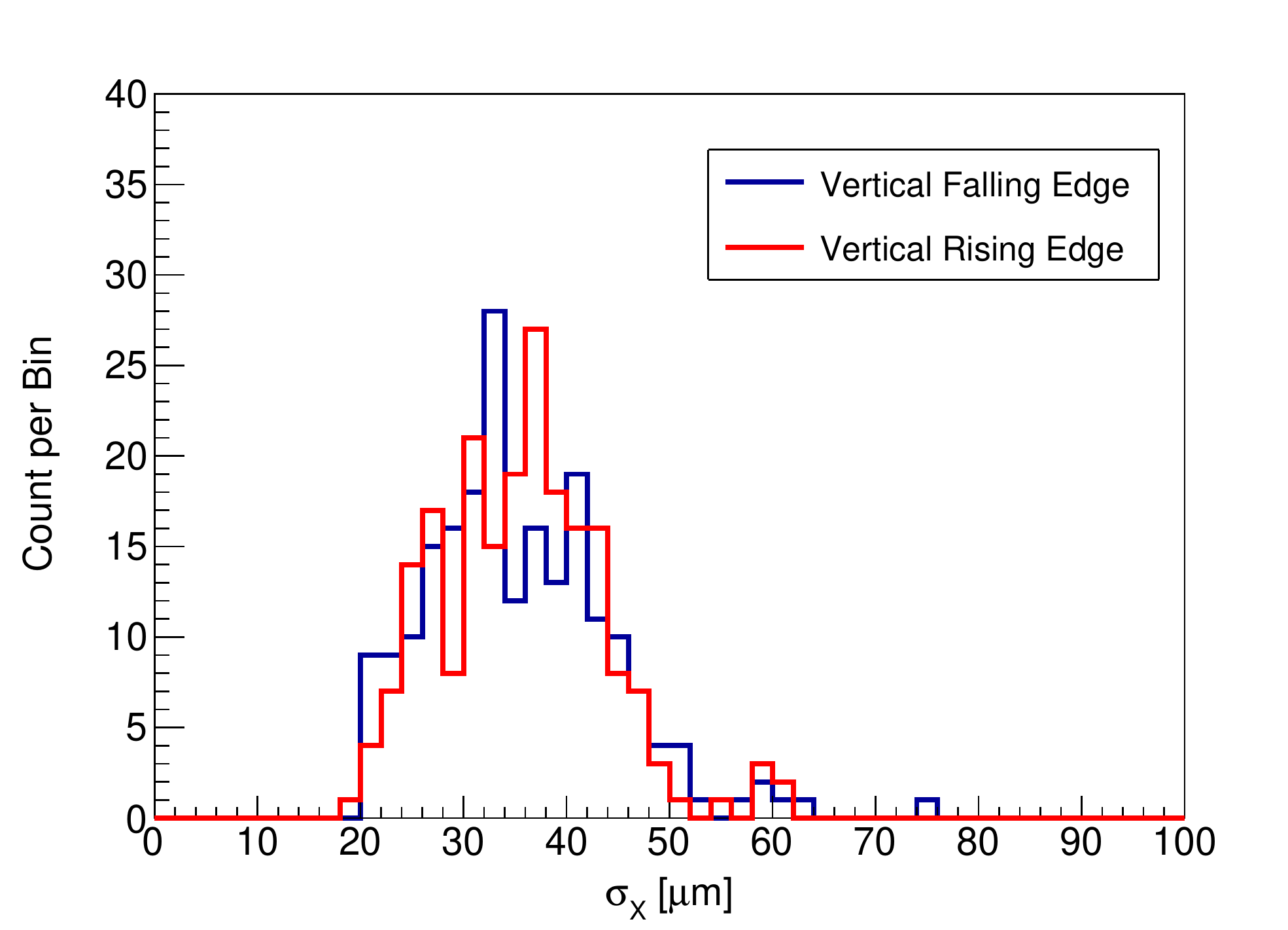}}
\caption{Distributions of Gaussian smearing widths of falling edges and rising edges. \label{Fig_MCPResStat}}
\end{figure}

The resolution of the MCP position response is reflected by the $\sigma$ parameters in Equation~\ref{Eq_MCPGridResponse}. $\sigma_{1}$ is the standard deviation of the Gaussian smearing corresponding to the falling edge, and $\sigma_{2}$ is the one corresponding to the rising edge. Distributions of fitted values of the $\sigma$ parameters for the intermediate points on horizontal and vertical segments are plotted in Figure~\ref{Fig_MCPResStat}. All these distributions have a similar shape with mean 36~$\mu$m and RMS 8~$\mu$m. The RMSs of these distributions are also comparable to the fit uncertainties of $\sigma_{1}$ and $\sigma_{2}$ in Equation~\ref{Eq_MCPGridResponse}, which are $\sim$6~$\mu$m. This indicates that the widths of the distributions in Figure~\ref{Fig_MCPResStat} are mainly due to the statistical uncertainties of fits.

\begin{figure}[h!]
\centering
\includegraphics[width=0.5\textwidth]{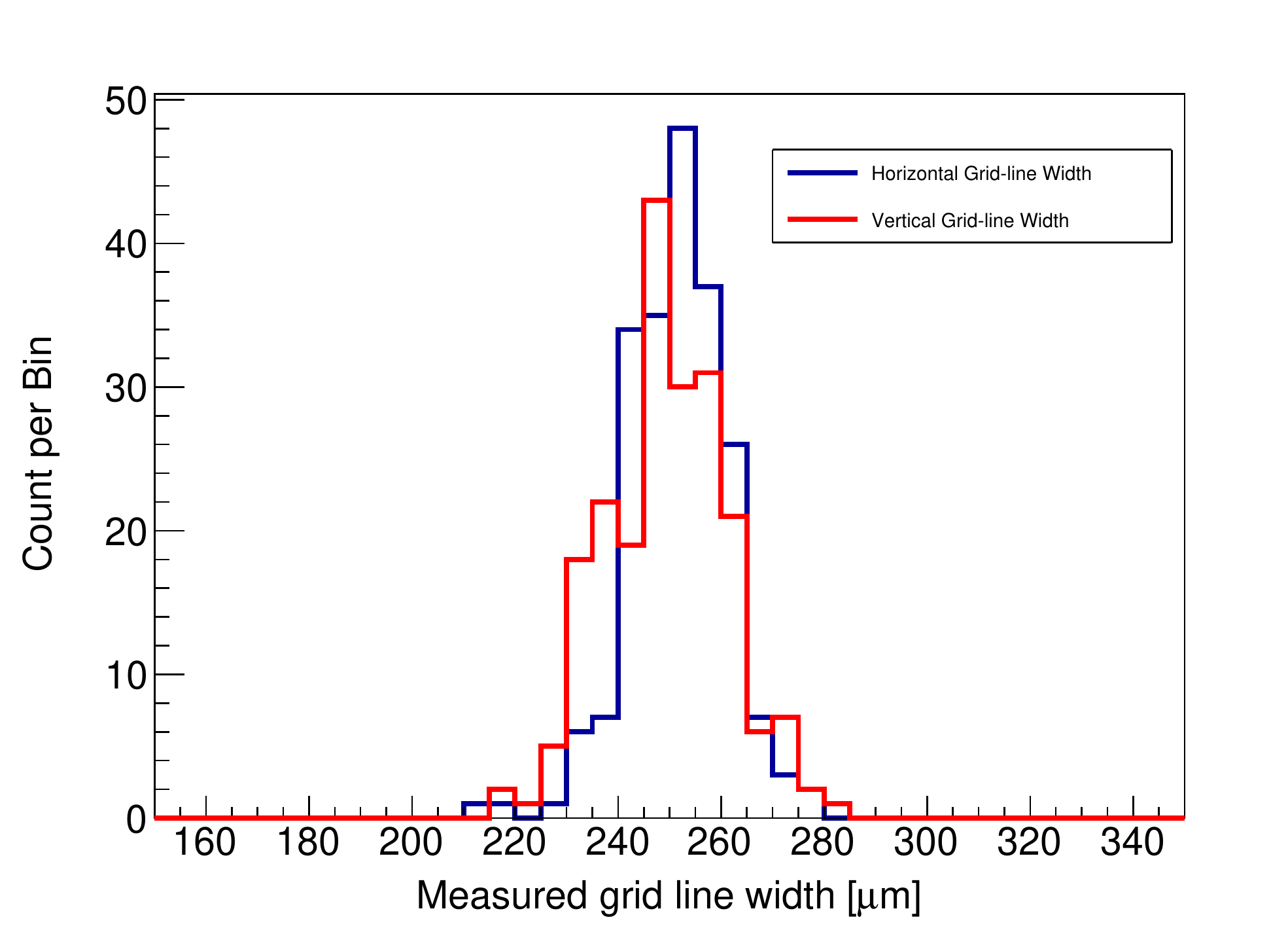}
\caption{Distributions of the widths of grid line segments. \label{Fig_MCPGridWidth}}
\end{figure}

Furthermore, the fitted value of $\mu_{2}-\mu_{1}$ corresponds to the width of the grid lines. The distributions of the widths of horizontal and vertical grid line segments are shown in Figure~\ref{Fig_MCPGridWidth}. The mean values of both the horizontal and vertical grid line widths are consistent with their physical width of 250~$\mu$m. We also investigated the stability of the calibration, and no significant drift was observed within two days.

\subsection{Summary}
\label{SubSec_CalSummary}

The position calibration scheme described in this section improves the MCP position readout accuracy significantly. After the event-by-event  correction, the accuracy\footnote{Because the grid line positions are determined by averaging over contributions from many channels, the accuracy of the grid line center determination can be better than the size of the micro-channels.} (mean-absolute-deviation) of the MCP position responses is 8~$\mu$m, and the RMS resolution (mean-$\sigma$-parameter) is 36~$\mu$m which corresponds to 85~$\mu$m FWHM. To achieve these calibration goals, the number of detected $\alpha$ particles needs to be at least $1.4\times10^{7}$. Compared to conventional calibration methods used so far, our calibration scheme improves the accuracy of the MCP position response by about one order of magnitude. Moreover, the excellent FWHM spatial resolution of 85~$\mu$m could be determined precisely, which is crucial for the Monte Carlo simulation of the detector response function in precision experiments such as the $^{6}$He $\beta-\nu$ angular correlation measurement \cite{Hong2016}. Since the spatial resolution is at the same scale of the channel-to-channel distance, using MCPs with smaller and denser micro-channels can potentially further improve the position resolution. 

So far all results we presented are based on the fact that the data run and the calibration run are under the same condition and the same type of particle is detected in both runs. We will show in the following section that the calibration parameters, can differ when different sources of particles are used or the $\alpha$-source position is different.

\section{Discussion}
\label{Sec_Discussion}

\subsection{Calibrations using different particle sources}
\label{SubSec_OtherSources}

Besides $\alpha$ sources, the position response of the MCP can also be calibrated using other ion sources. It is necessary to ensure that the MCP position response to $\alpha$ particles and to the particles detected in the target experiment are the same so that the high-accuracy calibration is still valid. For the experiment dedicated to the measurement of the $\beta-\nu$ angular correlation in $^{6}$He decay, the MCP is used to detect $^{6}$Li ions. Therefore, we also use the $^{6}$Li recoil ions from the $^{6}$He decays to illuminate the MCP and calibrate its position response. High-intensity $^{6}$Li ions are achieved by flushing the $^{6}$He-trap chamber with non-trapped $^{6}$He atoms and turning on the ion-accelerating field. The recoil $^{6}$Li ions are thus accelerated towards the detector with an average energy of $\approx$16~keV to trigger the MCP efficiently. The $^{6}$Li ions are blocked by the mask and make the grid line shadows on the MCP image clearly visible. However, the contrast of MCP image is not as good as that in the calibration run using $\alpha$ particles, because most of the $\beta$ particles from $^{6}$He decays can penetrate the calibration mask and trigger the MCP. Although a small fraction of the $\beta$ particles can be blocked by the mask, our study shows that the contribution from $\beta$ particles to the grid line shadow is negligible compared to the contribution from $^{6}$Li recoil ions. Therefore, the position spectrum generated by $\beta$ particles from $^{6}$He decays can be treated as a constant background at each fit region when determining the grid line positions.

\begin{figure}[h!]
\centering
\subfloat[Distribution of absolute deviations]{\includegraphics[width=0.49\textwidth]{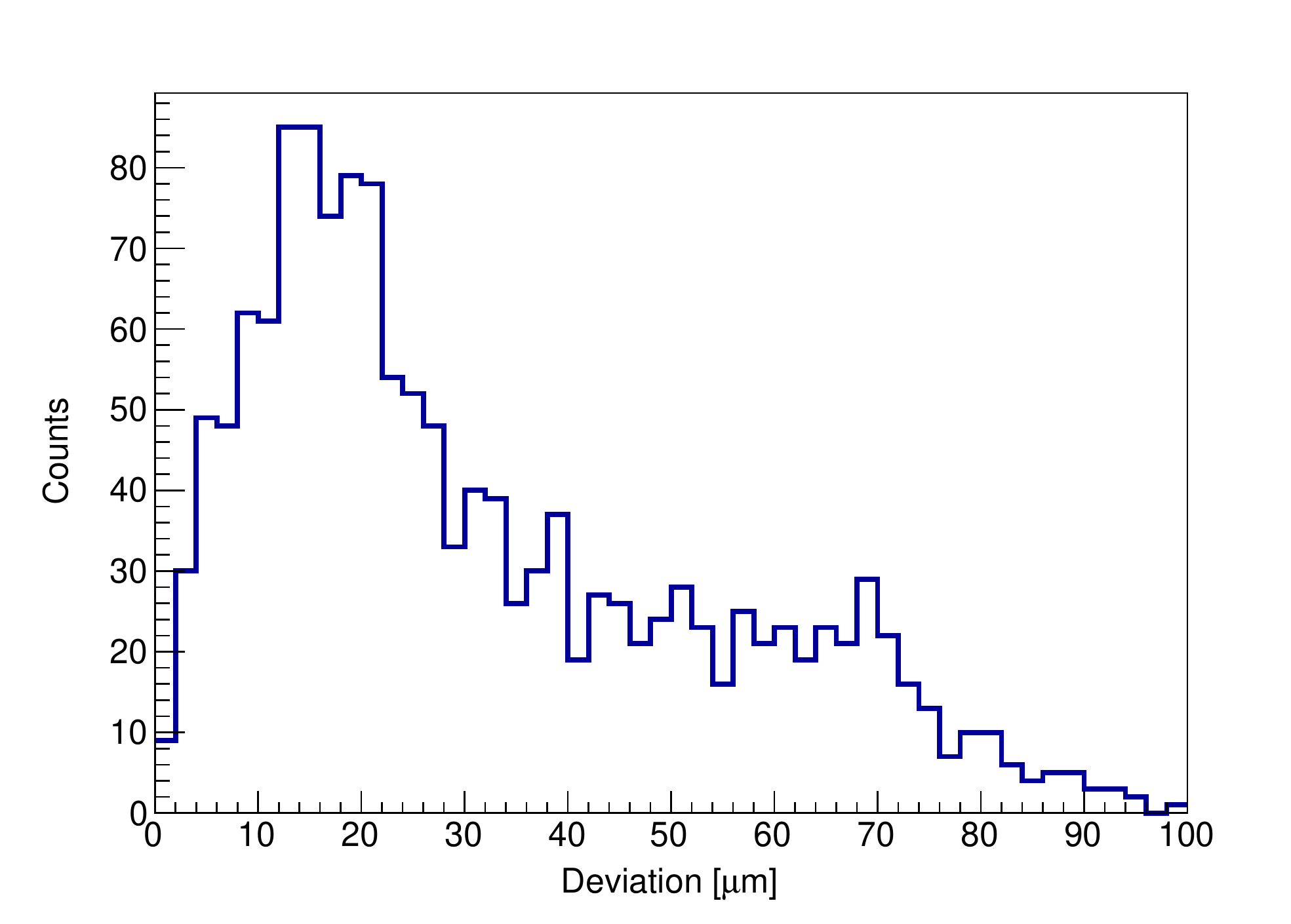}}~~
\subfloat[Distribution of RMS resolutions]{\includegraphics[width=0.49\textwidth]{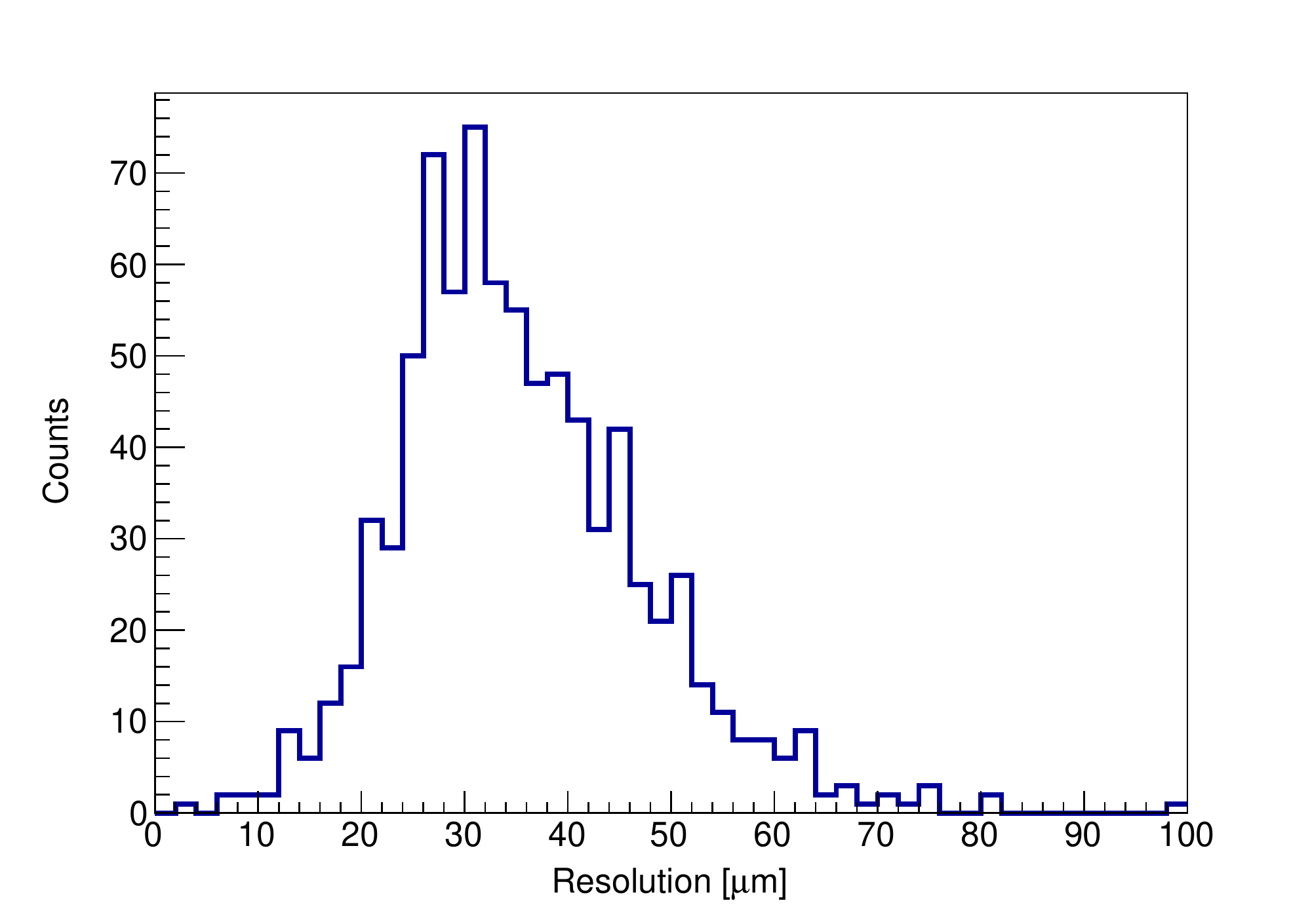}}
\caption{Distributions of deviations and resolutions for the MCP position spectrum of the non-trapped $^{6}$He source calibrated based on the run with $\alpha$-particles. \label{Fig_MCPResStatLi6_AlpahCal}}
\end{figure}

\begin{figure}[h!]
\centering
\includegraphics[width=0.49\textwidth]{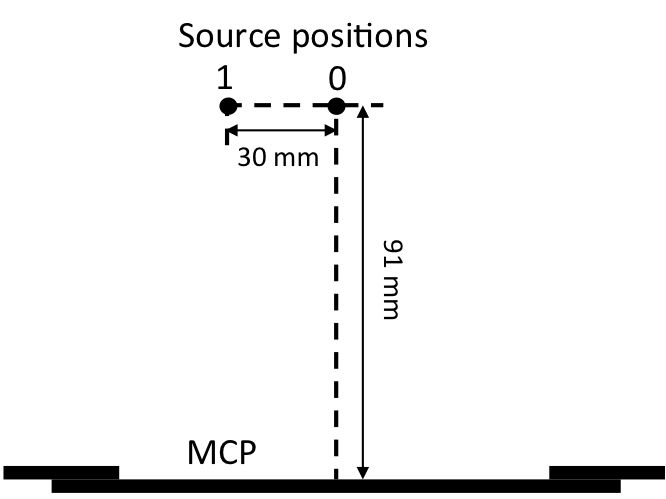}
\caption{Source positions for studying the MCP position calibration. Position-0 is for the MCP position calibration procedures described in Section~\ref{Sec_PosCal}. Position-1 is off-centered by 30~mm for studying the deviation patterns when the run with the $^{241}$Am source is calibrated by the run using non-trapped $^{6}$He atoms. \label{Fig_SourcePosition}}
\end{figure}

\begin{figure}[h!]
\centering
\subfloat[$^{241}$Am source above the MCP center]{\includegraphics[width=0.49\textwidth]{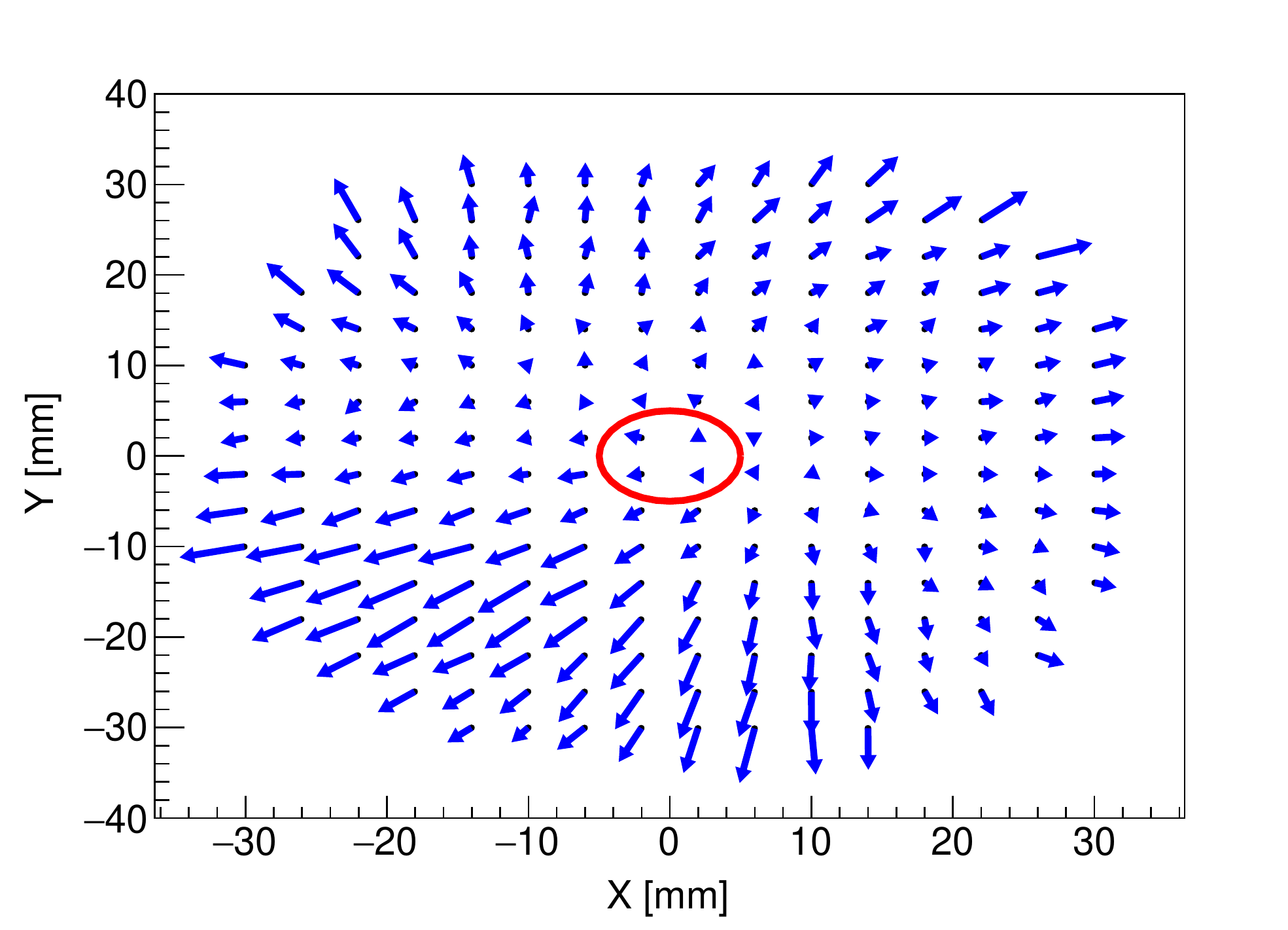}\label{Fig_AlphaCalDeviationPattern1}}~~
\subfloat[$^{241}$Am source off-centered by 30~mm]{\includegraphics[width=0.49\textwidth]{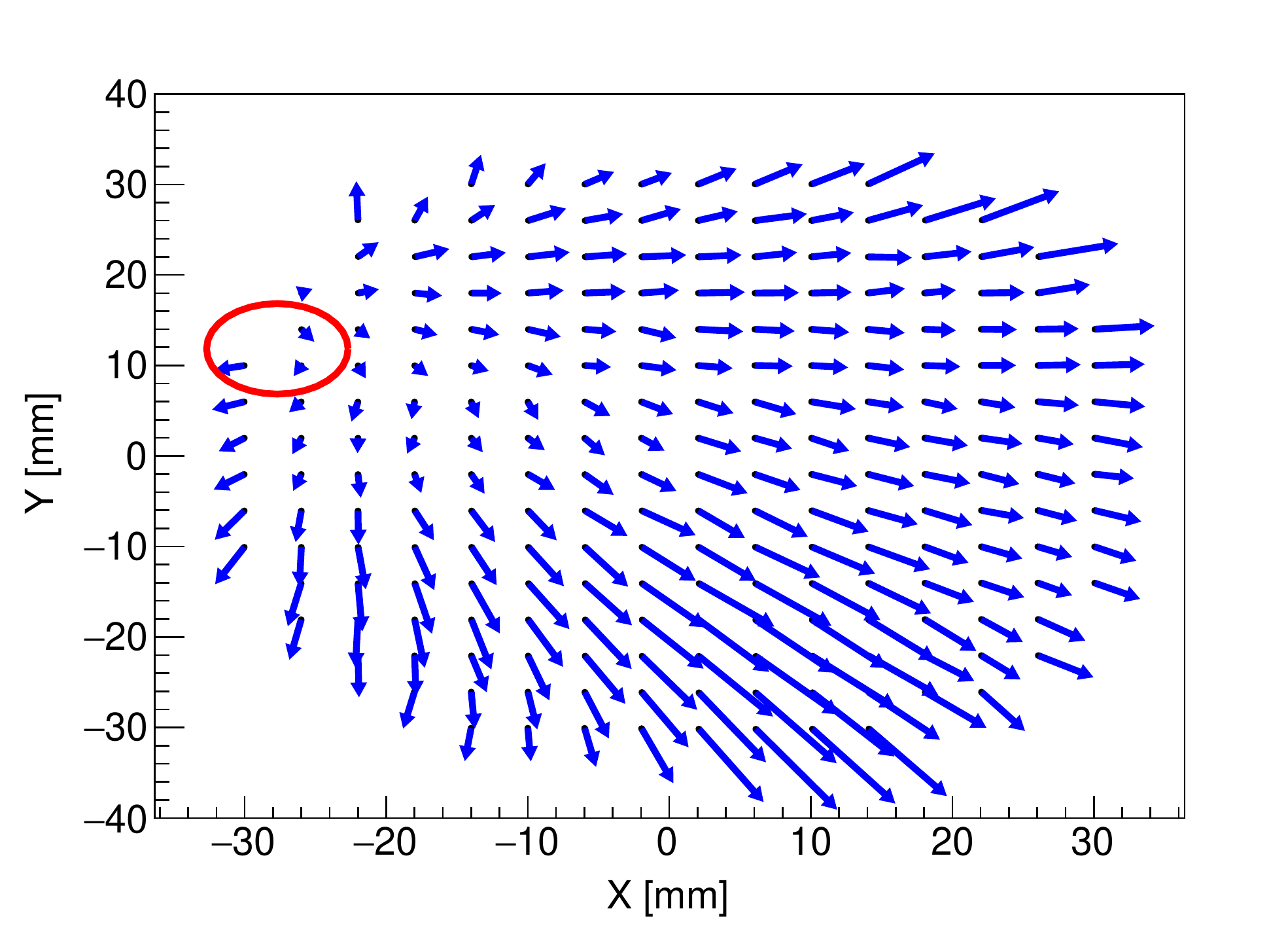}\label{Fig_AlphaCalDeviationPattern2}}
\caption{Patterns of grid line crossing deviations for two $\alpha$-particle runs calibrated using the calibration parameters obtained from a run using non-trapped $^{6}$He atoms. Each arrow in these two subfigures starts from the physical position of each grid line crossing and points to the direction of deviation. The length of each arrow is proportional to each absolute deviation. The $^{241}$Am source positions (projected onto the MCP) are indicated by the red circles. The micro-channels are inclined in the azimuthal orientation roughly 45$^{\circ}$ counterclockwise from the X axis.\label{Fig_AlphaCalDeviationPattern}}
\end{figure}

The calibration scheme described in Section~\ref{Sec_PosCal} is still valid for calibration runs using $^{6}$Li ions. We achieved the same accuracy (8~$\mu$m) and resolution (85~$\mu$m FWHM) as those obtained using $\alpha$ particles. However, the fitted value of the background parameter $B$ in Equation~\ref{Eq_MCPGridResponse} is $\approx$1/6 of the fitted value of $N_{1}$ and $N_{2}$ due to the $\beta$ particles, while for $\alpha$ particles $B$ is close to zero. Furthermore, we inspected the consistency between the calibration using the $^{241}$Am source and the calibration using the non-trapped $^{6}$He atoms. We first constructed the correction functions based on a run with the $^{241}$Am source, and then applied these correction functions to a run with non-trapped $^{6}$He atoms. We analyzed the deviations of the calibration points (after the position corrections) from their physical positions and the MCP resolution. The results are plotted in Figure~\ref{Fig_MCPResStatLi6_AlpahCal}. The distribution of the resolution is broader than that for cases where the calibration run uses the same species of particles, but the mean RMS resolution is still 36~$\mu$m. However, the mean absolute deviation becomes 32~$\mu$m and its distribution is much broader and extends to 100~$\mu$m. To understand the cause of those large deviations, we studied the {\em pattern of deviations} of the grid line crossings as shown in Figure~\ref{Fig_AlphaCalDeviationPattern}. Figure~\ref{Fig_AlphaCalDeviationPattern1} and Figure~\ref{Fig_AlphaCalDeviationPattern2} correspond respectively to two runs with the $^{241}$Am source placed above the center of the MCP (Position-0 in Figure~\ref{Fig_SourcePosition}) and off-centered by 30~mm (Position-1 in Figure~\ref{Fig_SourcePosition}). The position correction functions for these two runs are based on the same calibration run using non-trapped $^{6}$He atoms. As shown in these figures, the deviations are all diverging and the center of the divergence approximately moves along with the position of the $^{241}$Am source. On the other hand, according to the geometry of the micro-channels $\alpha$ particles with 5.5~MeV energy can penetrate the walls of the micro-channels once, and even twice for large incident angles. When an $\alpha$ particle penetrates more than one micro-channels, the reconstructed hit position consequently deviates towards the direction of the incoming $\alpha$ particle from the hit position on the MCP front surface. Therefore, we attribute the large deviations to the multi-channel penetration of the $\alpha$ particles. 

\begin{figure}[h!]
  \centering
  \includegraphics[width=0.7\linewidth]{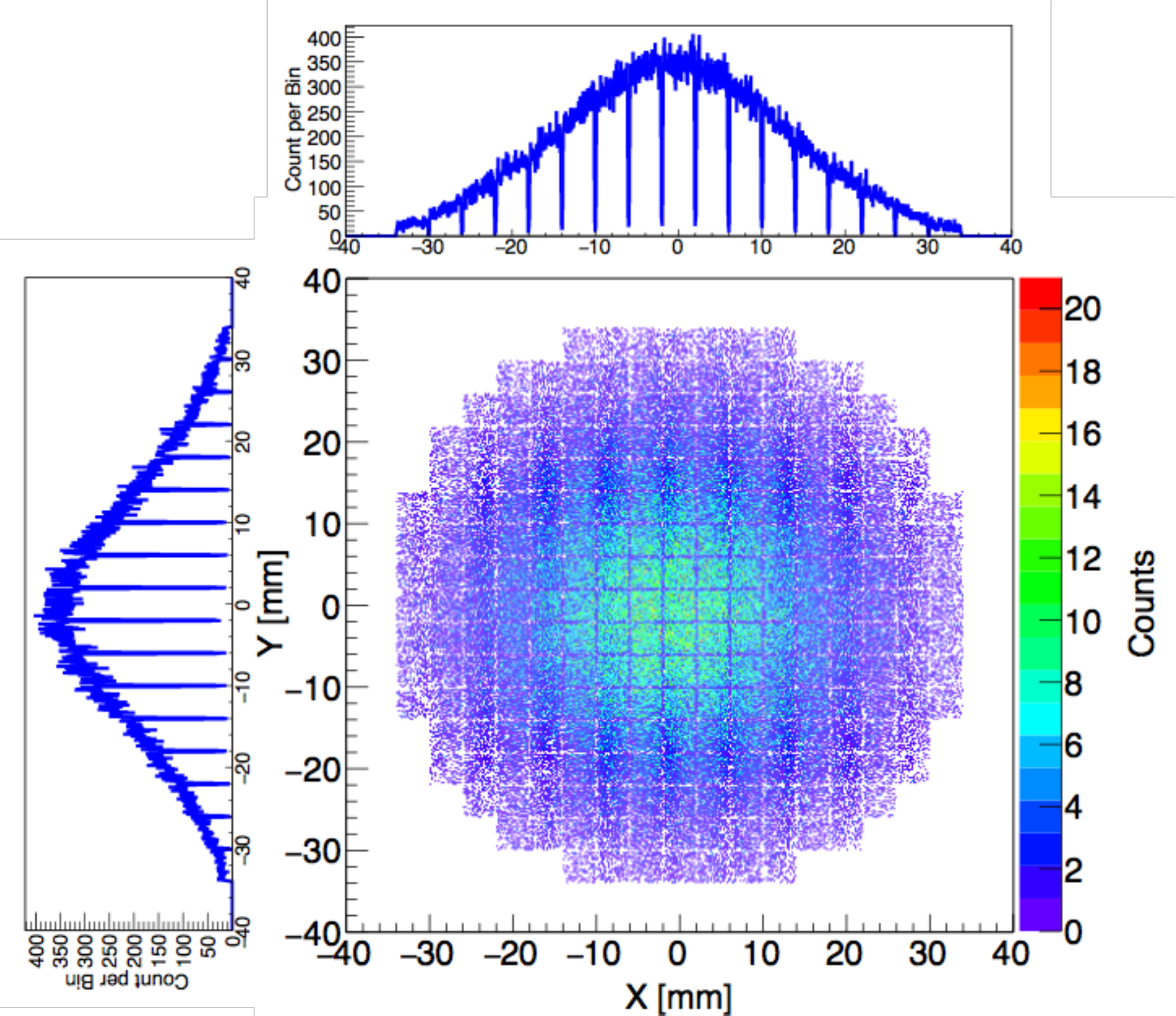}
\caption{MCP image of the $a_{\beta\nu}$ measurement. Projections of the image onto X and Y axis are shown in the top and left panels. The measured dip positions corresponding to the grid lines deviate from their physical positions by no more than 20~$\mu$m, which is consistent with the absolute-deviation distribution in Figure~\ref{Fig_MCPCorrectedAll}. \label{Fig_TrapHe6MCPImage}}
\end{figure}

Because the low-energy $^{6}$Li ions cannot penetrate through the walls of micro-channels, we choose to use the non-trapped $^{6}$He source to calibrate the MCP position response during the $^{6}$He decay $\beta-\nu$ correlation experiment. In this experiment, the $^{6}$He atoms are trapped 91~mm above the MCP center as shown in Figure~\ref{Fig_DetectorSysSchematic}, and the FWHM of the trap is $\sim$1~mm. The MCP is triggered in coincidence with the $\beta$ telescope, and a TOF cut is applied so that the data set contains only events generated by $^{6}$Li ions emitted by the decays inside the trap. We checked whether the reconstructed grid line positions on the calibrated MCP image shown in Figure~\ref{Fig_TrapHe6MCPImage} are consistent with their physical positions. Projection histograms of the MCP image are constructed, and the dip positions in these histograms correspond to the reconstructed grid line positions. We confirmed that the dip positions deviate from their physical positions by no more than 20~$\mu$m, which is consistent with the accuracy of the calibration.

\subsection{Impact on the $\beta-\nu$ angular correlation measurement of the $^{6}$He decay}
\label{SubSec_Impacts}

For the measurement of the $\beta-\nu$ angular correlation coefficient $a_{\beta\nu}$ in $^{6}$He decay, we aim at a relative precision of $\sim$1\% in the development stage of this experiment and then at an order of 0.1\% in the coming years. The sensitivity of the $a_{\beta\nu}$ extraction to the position of the boundaries of the fiducial area ($r_{cut}$) is approximately $\Delta a_{\beta\nu}=$~1.7\% per mm change in $r_{cut}$. Therefore, the 8~$\mu$m accuracy of the MCP position response corresponds to a 0.014\% systematic uncertainty of $a_{\beta\nu}$ which is negligible for the present experiment goals and in the near future. 

\section{Conclusions}
\label{Conclusions}

We developed a high-accuracy and high-resolution position calibration scheme for MCP detectors. It uses a 90\% open calibration mask permanently mounted on the detector. Software was developed to determine the grid line positions on the MCP image with high accuracy so that the adequate parameters for the event-by-event position correction could also be determined with high accuracy. An accuracy of 8~$\mu$m and a FWHM resolution of 85~$\mu$m were achieved, improving by close to $1$ order of magnitude the final accuracy of the image reconstruction compared to the conventional calibration schemes. Due to the high accuracy position response of the MCP, the related systematic uncertainty in the $a_{\beta\nu}$ measurement of the $^{6}$He decay is negligible. At this level of accuracy, systematic differences between different particles used for calibration become apparent and need to be taken into account. Stability of the calibration also needs to be monitored throughout the experiment.

\section{Acknowledgments}
This work is supported by the Department of Energy, Office of Nuclear Physics, under contract numbers DE-AC02-06CH11357 and DE-FG02-97ER41020.

\bibliographystyle{elsart-num}
\bibliography{bibliography}

\end{document}